\documentclass{article}
\pdfoutput=1

\usepackage[usenames,dvipsnames]{xcolor}
\usepackage{amsmath,amssymb,array,marvosym,multirow,rotating,pgf,tikz,stmaryrd,chemarrow,colortbl,listings}
\usepackage{amsthm}
\theoremstyle{definition}
\newtheorem{definition}{Definition}

\theoremstyle{theorem}
\newtheorem{proposition}{Proposition}
\newtheorem{lemma}{Lemma}
\newtheorem{theorem}{Theorem}
\newtheorem{corollary}{Corollary}
\theoremstyle{definition}

\newtheorem{fact}{Fact}

\usetikzlibrary{arrows,automata}

\usepackage{hyperref}

\newcommand{\mbbb}{-!!!}
\newcommand{\mbb}{-!!}
\newcommand{\mb}{-!}
\newcommand{\m}{-}
\newcommand{\n}{0}
\newcommand{\p}{+}
\newcommand{\pb}{+!}
\newcommand{\pbb}{+!!}
\newcommand{\pbbb}{+!!!}
\newcommand{\pmlb}{\pm_{!}}
\newcommand{\pmub}{\pm^{!}}

\newcommand{\F}[2]{\mbox{$\textsf{#1}#2$}}

\newcommand{\signatures}{\mathbb{S}}
\newcommand{\factors}{\mathbb{F}}

\newcommand{\SPP}{\textrm{SPP}}

\newcommand{\SPPs}{\mathrm{SPP^{+}}}
\newcommand{\SPPso}{\mathcal{STR}}

\newcommand{\conc}{\star}

\newcommand{\leftI}{\mathrm{p}}

\newcommand{\atoms}{\mathbb{A}}
\newcommand{\limp}{\rightarrow}
\newcommand{\true}{\top}
\newcommand{\false}{\bot}
\newcommand{\statements}{\mathcal{L}}

\newcommand{\Clo}[2]{\mathrm{Cl}_{#1}^{#2}}

\newcommand{\rightG}[1]{{#1^{\triangleright}}}
\newcommand{\leftG}[1]{{#1^{\triangleleft}}}
\newcommand{\rightleftG}[1]{{#1^{\triangleright\triangleleft}}}
\newcommand{\leftrightG}[1]{{#1^{\triangleleft\triangleright}}}

\newcommand{\invs}{\mathcal{I}}

\newcommand{\id}[1]{\mathrm{id}_{#1}}

\newcolumntype{D}{>{\centering\arraybackslash}p{3.5ex}}
\newcolumntype{C}{>{\centering\arraybackslash}p{0.26\textwidth}}
\newcolumntype{M}{>{\centering\arraybackslash}p{2.25ex}}

\begin{document}
\title{Computer-Aided Discovery and\\ Categorisation of Personality Axioms}
\author{Simon Kramer\\[\jot]
		\texttt{simon.kramer@a3.epfl.ch}}
\maketitle
\begin{abstract}
	We propose a computer-algebraic, order-theoretic framework based on 
		intuitionistic logic for 
			the computer-aided discovery of personality axioms from personality-test data and 
			their mathematical categorisation into formal personality theories in the spirit of 
				F.~Klein's \emph{Erlanger Programm} for geometrical theories.
	As a result,
		formal personality theories can be 
			automatically generated, 
			diagrammatically visualised, and 
			mathematically characterised in terms of 
				categories of invariant-preserving transformations in the sense of Klein and category theory.
	Our personality theories and categories are induced by  
		implicational invariants 
			that are ground instances of intuitionistic implication, which
				we postulate as axioms.
	In our mindset, 
		the essence of personality, and thus mental health and illness, is its invariance.
	The truth of these axioms is algorithmically extracted from 
		histories of partially-ordered, symbolic data of observed behaviour.
	The personality-test data and the personality theories are related by a Galois-connection in our framework.
	As data format, we adopt the format of the symbolic values generated by 
		the Szondi-test, a personality test based on 
			L.~Szondi's unifying, depth-psychological theory of fate analysis.
	
	\smallskip
	
\end{abstract}

\section{Introduction}
In 1872,
	Felix Klein, 
		full professor of mathematics at the University of Erlangen at age 23, 
			presented his influential \emph{Erlanger Programm} \cite{Klein:ErlangerProgramm,Klein:Geometry} on 
				the classification and characterisation of geometrical theories by
					means of group theory.
That is,
	Klein put forward the thesis that 
		every geometrical theory could be characterised by 
			an associated group of geometrical transformations that 
				would leave invariant the essential properties of 
					the geometrical objects of that theory.
These essential properties are captured by the axioms that define the theory.
As a result,
	geometrical theories could be classified in terms of
		their associated transformation groups.
According to \cite{ErlangerProgramm},
	Klein's \emph{Erlanger Programm} 
		``is regarded as one of the most influential works in the history of geometry, and 
			more generally mathematics, during the half-century after its publication in 1872.''\footnote{%
				We add that in physics, the existence of certain transformation groups for 
					mechanics and electromagnetism led Albert Einstein to discover his theory of special relativity.}

In this paper and in the spirit of 
	the \emph{Erlanger Programm} for geometrical theories,
	we propose a computer-algebraic,\footnote{in the sense of \emph{symbolic} as opposed to numeric computation} 
		order-theoretic framework based on 
		intuitionistic logic \cite{sep-logic-intuitionistic} for 
			the computer-aided discovery of personality axioms from personality-test data and 
			their mathematical categorisation into formal personality theories.
Each one of the resulting intuitionistic personality theories is 
	an (order-theoretic) \emph{prime filter} \cite{DaveyPriestley} in our framework.
As our contribution, 
	formal personality theories can be 
		automatically generated, 
		diagrammatically visualised, and 
		mathematically characterised in terms of 
			categories of invariant-preserving transformations in the sense of Klein and category theory \cite{sep-category-theory}.
That is,
	inspired by and in analogy with Klein,
		we put forward the thesis that 
			every personality theory can be characterised by
				an associated category of personality transformations that 
					leave invariant the essential properties of 
						``the personality objects''---the people, 
							represented by their personality-test data---of that theory.

An important difference in our psychological context of personality theories to 
	Klein's geometrical context is that
		actually no formal personality theory in the strict axiomatic sense exists, whereas 
			Klein could characterise a variety of \emph{existing,} axiomatic theories of geometry.
Before being able to categorise personality theories,
	we thus must first formally \emph{define} them.
As said,
	we shall do so by 
		discovering their defining axioms from 
			personality-test data with the aid of computers.
Our personality theories and categories are then automatically induced by  
	implicational invariants that hold throughout that test data and 
		that are ground instances of intuitionistic implication, which
			we postulate as axioms.
In our mindset, 
	the essence of personality, and thus mental health and illness, is its invariance.
So for every person, represented by her personality-test result $P$---the data---we 
	automatically generate her associated  
		\begin{enumerate}
			\item personality \emph{theory} $\leftG{\{P\}}$ of 
					simple implicational invariants and 
			\item personality \emph{category} $\mathbf{T}_{\leftG{\{P\}}}$ of 
					theory-preserving transformations.
		\end{enumerate}
(We are actually able to carry out this construction for whole \emph{sets} of personality-test results, either 
		of different people or of one and the same person.)
More precisely,
	the truth of these axioms is algorithmically extracted from 
		histories of partially-ordered, symbolic data of the person's observed test behaviour.
Our axioms have an implicational form in order to 
	conform with the standard of Hilbert-style axiomatisations \cite{WhatIsALogicalSystem}, which 
		in our order-theoretic framework can be cast as a simple closure operator.
Another difference in our context is that 
	contrary to Klein, who
		worked with transformation groups,  
			we work with more general transformation \emph{monoids,} and 
				thus transformation categories.
The reason is that 
	contrary to Klein's geometrical context, in which
		transformations are invertible, 
			transformations in the psychological context need not be invertible.

In our order-theoretic framework, 
	personality-test data and personality theories are related by a Galois-connection 
		$(\,\rightG{}\,,\leftG{}\,)$ \cite[Chapter~7]{DaveyPriestley}.
As data format, 
	we adopt---without loss of generality---the format of the symbolic values, 
		called \emph{Szondi personality profiles (SPPs),} generated by 
			the Szondi-test \cite{Szondi:ETD:Band1}, a personality test based on 
				L.~Szondi's unifying, depth-psychological theory of fate analysis \cite{Szondi:IchAnalyse}.
An SPP can be conceived as a tuple of eight, 
		so-called \emph{signed factors} whose signatures can in turn take twelve values.
We stress that 
	our framework is 
		independent of any personality test.
It simply operates on 
	the result values that such tests generate.
Our choice of the result values of the Szondi-test is motivated by the fact that 
	SPPs just happen to have a finer structure than 
		other personality-test values that we are aware of, and 
			so are perhaps best suited to play 
				the illustrative role for which we have chosen them here.
(See also \cite{arXiv:1403.2000v1}.)

The remaining part of this paper is structured as follows:
in Section~\ref{section:Discovery}, we present 
	the part of our framework for the computer-aided discovery of personality axioms from
		personality-test data, and 
in Section~\ref{section:Categorisation}, 
	the part for their mathematical categorisation into formal personality theories.

\section{Axiom discovery}\label{section:Discovery}
In this section,
	we present 
		the part of our framework for the computer-aided discovery of personality axioms from
			personality-test data.
This is the data-mining and the logical and visual data-analytics part of our contribution.

We start with defining the format of the data on which 
	we perform our data-mining and data-analytics operations.
As announced,
	it is the format of the symbolic values, 
		called \emph{Szondi personality profiles (SPPs),} generated by 
			the Szondi-test \cite{Szondi:ETD:Band1}.
We operate on finite sequences thereof.
In diagnostic practice,
	these test-result sequences are usually composed of 10 SPPs \cite{SzondiTestWebApp}.

\begin{definition}[The Szondi-Test Result Space]\label{definition:SPP}
Let us consider the Hasse-diagram \cite{DaveyPriestley} in Figure~\ref{figure:SzondiSignatures} 
\begin{figure}[t]
\centering
\caption{Hasse-diagram of Szondi's signatures}
\medskip
\fbox{\begin{tikzpicture}
	\node (pbbb) at (0,4) {$+!!!$};
	\node (pbb) at (0,3) {$+!!$};
	\node (pb) at (0,2) {$+!$};
	\node (p) at (0,1) {$+$};
	\node (n) at (0,0) {$0$};
	\node (m) at (0,-1) {$-$};
	\node (mb) at (0,-2) {$-!$};
	\node (mbb) at (0,-3) {$-!!$};
	\node (mbbb) at (0,-4) {$-!!!$};
	\draw (mbbb) -- (mbb) -- (mb) -- (m) -- (n) -- (p) -- (pb) -- (pbb) -- (pbbb);	
 	\node (pmub) at (1,1) {$\pm^{!}$};  
	\node (pm) at (1,0) {$\pm$};
	\node (pmlb) at (1,-1) {$\pm_{!}$};
	\draw (pmlb) -- (pm) -- (pmub);
\end{tikzpicture}}
\label{figure:SzondiSignatures}
\end{figure}
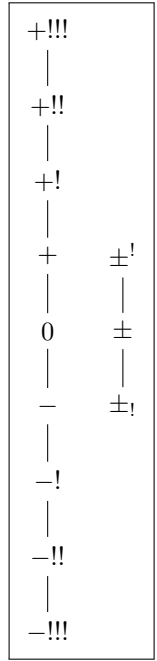
of the partially ordered set of \emph{Szondi's twelve signatures} \cite{Szondi:ETD:Band1} of 
human reactions, which are:
\begin{itemize}
	\item approval: from strong $+!!!$\,, $+!!$\,, and $+!$ to weak $+$\,; 
	\item indifference/neutrality: $0$\,; 
	\item rejection: from weak $-$\,, $-!$\,, and $-!!$ to strong $-!!!$\,; and 
	\item ambivalence: $\pm^{!}$ (approval bias), $\pm$ (no bias), and $\pm_{!}$ (rejection bias).
\end{itemize}
(Szondi calls the exclamation marks in his signatures \emph{quanta.})

Further let us call this set of signatures $\mathbb{S}$, that is,
		$$\signatures:=\{\,\mbbb,\mbb,\mb,\m,\n,\p,\pb,\pbb,\pbbb,\pmlb,\pm,\pmub\,\}.$$	
	
Now let us consider \emph{Szondi's eight factors and four vectors} of 
	human personality \cite{Szondi:ETD:Band1} as summarised in Table~\ref{table:SzondiFactors}.
	\begin{table}[t]
		\centering
		\caption{Szondi's factors and vectors}
		\medskip
		\begin{tabular}{|c|c||C|C|}
			\hline
			\multirow{2}{12.5ex}{\centering \textbf{Vector}} & \multirow{2}{7.5ex}{\textbf{Factor}} & \multicolumn{2}{c|}{\textbf{Signature}}\\
			\cline{3-4}
			&& $+$ & $-$\\
			\hline
			\hline
			\multirow{2}{12.5ex}{\centering \textsf{S} (Id)} & \textsf{h} (love) & physical love & platonic love\\
			\cline{2-4}
			& \textsf{s} (attitude) & (proactive) activity & (receptive) passivity\\
			\hline
			\multirow{2}{12.5ex}{\centering \textsf{P}\\[-\jot] (Super-Ego)} & \textsf{e} (ethics) & ethical behaviour & unethical behaviour\\
			\cline{2-4}
			& \textsf{hy} (morality) & immoral behaviour & moral behaviour\\
			\hline
			\multirow{2}{12.5ex}{\centering \textsf{Sch} (Ego)} & \textsf{k} (having) & having more & having less\\
			\cline{2-4}
			& \textsf{p} (being) & being more & being less\\
			\hline			
			\multirow{2}{12.5ex}{\centering \textsf{C} (Id)} & \textsf{d} (relations) & unfaithfulness & faithfulness\\
			\cline{2-4}
			& \textsf{m} (bindings) & dependence & independence\\
			\hline		
		\end{tabular}
		\label{table:SzondiFactors}
	\end{table}
(Their names are of clinical origin and need not concern us here.)
And let us call the set of factors $\factors$, that is, 
	$$\factors:=\{\,\F{h}{},\F{s}{},\F{e}{},\F{hy}{},\F{k}{},\F{p}{},\F{d}{},\F{m}{}\,\}.$$

Then, 
	\begin{itemize}
		\item $\SPP:=\{\; \begin{array}[t]{@{}l@{}}
				((\F{h}{,s_{1}}), (\F{s}{,s_{2}}), (\F{e}{,s_{3}}), (\F{hy}{,s_{4}}), 
				 (\F{k}{,s_{5}}), (\F{p}{,s_{6}}), (\F{d}{,s_{7}}), (\F{m}{,s_{8}})) \mid\\ 
				 s_{1},\ldots,s_{8}\in\signatures\;\}
				 \end{array}$
				 
				is the set of Szondi's personality profiles; 
		\item $\langle\,\SPPs,\conc\,\rangle$ is the free semigroup on 
				the set $\SPPs$ of all finite sequences of SPPs with
					$\conc$ the (associative) concatenation operation on $\SPPs$ and 
				$$\begin{array}[t]{@{}r@{\ \ }c@{\ \ }l@{}}
				\SPPs &:=&\bigcup_{n\in\mathbb{N}\setminus\{0\}}\SPP^{n}\\[2\jot]
				\SPP^{1+n} &:=&\SPP^{1}\times\SPP^{n}\\[\jot]
				\SPP^{1} &:=&\SPP\,;
				\end{array}$$
		\item $\SPPso:=\langle\,\SPPs,\sqsubseteq\,\rangle$
				is our \emph{Szondi-Test Result Space,} where
					the suffix partial order $\sqsubseteq$ on $\SPPs$ is defined such that
						for every $ P, P'\in\SPPs$,
							$ P\sqsubseteq P'$ if and only if
								$ P= P'$ or 
								there is $ P''\in\SPPs$ such that $ P= P''\conc P'$.

	\end{itemize}
\end{definition}
\noindent
As an example of an SPP,
	consider the \emph{norm profile} for the Szondi-test \cite{Szondi:ETD:Band1}:
		$$((\F{h}{,\p}), (\F{s}{,\p}), (\F{e}{,\m}), (\F{hy}{,\m}), 
				 (\F{k}{,\m}), (\F{p}{,\m}), (\F{d}{,\p}), (\F{m}{,\p}))$$
Spelled out, 
	the norm profile describes the personality of a human being who 
		approves of physical love,
		has a proactive attitude,
		has unethical but moral behaviour,
		wants to have and be less, and 
		is unfaithful and dependent.
		
Those SPP-sequences that 
	have been generated by 
		a Szondi-test(ee) are 
			our histories of 
				partially-ordered, symbolic data of 
					observed behaviour that we announced in the introduction.
Table~\ref{table:VGPSzondi} displays an example of such an SPP-sequence:
it is the so-called foreground profile of 
	a 49-year old, male physician and psycho-hygienist and 
		is composed of 10 subsequent SPPs \cite[Page~182--184]{Szondi:Triebpathologie:IA}.
	
\begin{fact}[Prefix closure of $\sqsubseteq$]\label{fact:conc}
	For every $P,P',P''\in\SPPs$, 
		$$\text{$P\sqsubseteq P'$ implies $P''\conc P\sqsubseteq P'$}$$
\end{fact}
\begin{proof}
	By inspection of definitions.
\end{proof}
	
\begin{table}[t]
\centering
\caption{A Szondi-test result (say $P$)}
\smallskip
\begin{tabular}{|D||D|D|D|D|D|D|D|D|}
\hline
\multirow{2}{3.2ex}{\centering \mbox{Nr.}} &\multicolumn{2}{c|}{\textsf{S}} & \multicolumn{2}{c|}{\textsf{P}} & \multicolumn{2}{c|}{\textsf{Sch}} & \multicolumn{2}{c|}{\textsf{C}}\\
\cline{2-9}
& \textsf{h} & \textsf{s} & \textsf{e} & \textsf{hy} & \textsf{k} & \textsf{p} & \textsf{d} & \textsf{m}\\
\hline
\hline
1 & $-$ & $0$ & $\pm$ & $\pm$ & $\pm$ & $\pm$ & $0$ & $+$\\
\hline
2 & $-$ & $0$ & $+$ & $\pm$ & $\pm$ & $+$ & $0$ & $+$\\
\hline
3 & $-$ & $-$ & $\pm$ & $\pm$ & $\pm$ & $+$ & $+$ & $\pm$\\
\hline
4 & $-$ & $-$ & $\pm$ & $+$ & $+$ & $+$ & $0$ & $+$\\
\hline
5 & $-$ & $0$ & $0$ & $+$ & $\pm$ & $\pm$ & $0$ & $+$\\
\hline
6 & $-$ & $0$ & $\pm$ & $\pm$ & $\pm$ & $\pm$ & $+$ & $\pm$\\
\hline
7 & $-$ & $0$ & $\pm$ & $\pm$ & $\pm$ & $+$ & $0$ & $+$\\
\hline
8 & $-$ & $-$ & $0$ & $\pm$ & $+$ & $+$ & $+$ & $\pm$\\
\hline
9 & $-$ & $0$ & $\pm$ & $\pm$ & $\pm$ & $\pm$ & $0$ & $+$\\
\hline
10 & $-$ & $0$ & $0$ & $\pm$ & $\pm$ & $+$ & $0$ & $+$\\
\hline
\end{tabular}
\label{table:VGPSzondi}
\end{table}
	
We continue to define the closure operator by which  
	we generate our intuitionistic personality theories from 
		personality-test data in the previously-defined format.
Our personality theories are intuitionistic, because
	such theories can be interpreted over 
		partially-ordered state spaces---such as our $\SPPso$---such that 
			a sentence is true in the current state by definition if and only if
				the sentence is true in all states that 
					are accessible from the current state by
						means of the partial order \cite{KripkeSemanticsIL,sep-logic-intuitionistic}.
In other words,
	the truth of such sentences is forward-invariant, which
		is precisely the property of sentences that we are looking for.
\begin{definition}[A closure operator for intuitionistic theories]\label{definition:AxiomsRules}
	Let 
		$$\atoms:=\{\,\F{h}{s_{1}}, \F{s}{s_{2}}, \F{e}{s_{3}}, \F{hy}{s_{4}}, \F{k}{s_{5}}, \F{p}{s_{6}}, \F{d}{s_{7}}, \F{m}{s_{8}}  \mid  
					s_{1},\ldots,s_{8}\in\signatures\,\}$$
	be our set of atomic statements, and 
	$$\statements(\atoms)\ni\phi::=A\mid\phi\land\phi\mid\phi\lor\phi\mid\neg\phi\mid\phi\limp\phi\quad\text{for $A\in\atoms$}$$
	our logical language over $\atoms$, that is, 
		the set of statements $\phi$ constructed from the atomic statements $A$ and
			the intuitionistic logical connectives 
				$\land$ (conjunction, pronounced ``and''),
				$\lor$ (disjunction, pronounced ``or''), 
				$\neg$ (negation, pronounced ``henceforth not''), and 
				$\limp$ (implication, pronounced ``whenever---then'').
	As usual, 
		we can macro-define 
			falsehood as $\false:=A\land\neg A$ and 
			truth as $\true:=\neg\false$\,.
				
Further let
\begin{flushleft}
	$\Gamma_{0} := \{$
	\vspace{-0.25\baselineskip}
			\begin{itemize}
				\item $\phi\limp(\phi'\limp\phi)$\\[-2\jot]
				\item $(\phi\limp(\phi'\limp\phi''))\limp((\phi\limp\phi')\limp(\phi\limp\phi''))$\\[-2\jot]
				\item $(\phi\land\phi')\limp\phi$\\[-2\jot]
				\item $(\phi\land\phi')\limp\phi'$\\[-2\jot]
				\item $\phi\limp(\phi'\limp(\phi\land\phi'))$\\[-2\jot]
				\item $\phi\limp(\phi\lor\phi')$\\[-2\jot]
				\item $\phi'\limp(\phi\lor\phi')$\\[-2\jot]
				\item $(\phi\limp\phi')\limp((\phi''\limp\phi')\limp((\phi\lor\phi'')\limp\phi'))$\\[-2\jot]
				\item $\false\limp\phi\;\}$\\[-2\jot]
			\end{itemize}
\end{flushleft}
be our (standard) set of intuitionistic \emph{axiom schemas.} 

	Then, $\Clo{}{}(\emptyset):=\bigcup_{n\in\mathbb{N}}\Clo{}{n}(\emptyset)$, where 
		for every $\Gamma\subseteq\statements(\atoms):$
		\begin{eqnarray*}
					\Clo{}{0}(\Gamma) &:=& \Gamma_{0}\cup\Gamma\\
					\Clo{}{n+1}(\Gamma) &:=& 
						\begin{array}[t]{@{}l@{}}
							\Clo{}{n}(\Gamma)\ \cup\\
							\{\,\phi'\mid\{\phi,\phi\limp\phi'\}\subseteq\Clo{}{n}(\Gamma)\,\}
								\quad\text{(\emph{modus ponens}, MP)}
						\end{array}
				\end{eqnarray*}
		We call $\Clo{}{}(\emptyset)$ our \emph{base theory,} and
		$\Clo{}{}(\Gamma)$ a \emph{$\Gamma$-theory} 
		for any $\Gamma\subseteq\statements(\atoms)$.
\end{definition}
\noindent
The following standard fact asserts that we have indeed defined a closure operator.
We merely state it as a reminder, because we shall use it in later proof developments.
The term $2^{\Gamma}_{\text{finite}}$ denotes the set of all finite subsets of the set $\Gamma$.
\begin{fact}\label{fact:ClosureOperator}
	The mapping ${\Clo{}{}}:2^{\statements(\atoms)}\rightarrow2^{\statements(\atoms)}$ is a \emph{standard consequence operator,} that is, 
				a \emph{substitution-invariant compact closure operator:}
					\begin{enumerate}
						\item $\Gamma\subseteq\Clo{}{}(\Gamma)$\quad(extensivity)
						\item if $\Gamma\subseteq\Gamma'$ then $\Clo{}{}(\Gamma)\subseteq\Clo{}{}(\Gamma')$\quad(monotonicity)
						\item $\Clo{}{}(\Clo{}{}(\Gamma))\subseteq\Clo{}{}(\Gamma)$\quad(idempotency)
						\item $\Clo{}{}(\Gamma)=\bigcup_{\Gamma'\in2^{\Gamma}_{\text{finite}}}\Clo{}{}(\Gamma')$\quad(compactness)
						\item $\sigma[\Clo{}{}(\Gamma)]\subseteq\Clo{}{}(\sigma[\Gamma])$\quad(substitution invariance), 
					\end{enumerate}
					where $\sigma$ designates an arbitrary propositional $\statements(\atoms)$-substitution.
\end{fact}
\begin{proof}
For (1) to (4),
	inspect the inductive definition of $\Clo{}{}$.
And (5) follows from our definitional use of axiom \emph{schemas}.\footnote{%
		Alternatively to axiom schemas,
		we could have used 
			axioms together with an
			additional substitution-rule set
				$\{\sigma[\phi]\mid\phi\in\Clo{}{n}(\Gamma)\}$
		in the definiens of $\Clo{}{n+1}(\Gamma)$.}
\end{proof}
\noindent
Note that in the sequel, 
	``:iff'' abbreviates ``by definition, if and only if,'' and 
	\begin{center}
		\begin{tabular}{rcl}
			$\Phi\vdash_{\Gamma}\phi$ &:iff& $\Phi\subseteq\Clo{}{}(\Gamma)$ implies $\phi\in\Clo{}{}(\Gamma)$\\[\jot]
			$\vdash_{\Gamma}\phi$ &:iff& $\emptyset\vdash_{\Gamma}\phi$\,.
		\end{tabular}
	\end{center}

\begin{table}[t]
	\caption{The diagram of $\invs(P)$ as extracted from the $P$ in Table~\ref{table:VGPSzondi}}
	{\smallskip\sf
	\resizebox{\textwidth}{!}{
\begin{tabular}{@{\;}rcc@{\quad}|cccc@{\quad}|cc}
&{\Large h}&{\Large s}&{\Large e}&{\Large hy}&{\Large k}&{\Large p}&{\Large d}&{\Large m}\\[2\jot]
{\Large h}&\renewcommand{\arraystretch}{1.5}
\begin{tabular}{|M|M|M|M|M|}
\hline
$\rightarrow$ &$0$&$+$&$-$&$\pm$\\
\hline
$0$&\cellcolor[gray]{0}&\cellcolor[gray]{0}&\cellcolor[gray]{0}&\cellcolor[gray]{0}\\
\hline
$+$&\cellcolor[gray]{0}&\cellcolor[gray]{0}&\cellcolor[gray]{0}&\cellcolor[gray]{0}\\
\hline
$-$&10&10&\cellcolor[gray]{0}&10\\
\hline
$\pm$&\cellcolor[gray]{0}&\cellcolor[gray]{0}&\cellcolor[gray]{0}&\cellcolor[gray]{0}\\
\hline
\end{tabular}\renewcommand{\arraystretch}{1.5}
&\renewcommand{\arraystretch}{1.5}
\begin{tabular}{|M|M|M|M|M|}
\hline
$\rightarrow$ &$0$&$+$&$-$&$\pm$\\
\hline
$0$&\cellcolor[gray]{0}&\cellcolor[gray]{0}&\cellcolor[gray]{0}&\cellcolor[gray]{0}\\
\hline
$+$&\cellcolor[gray]{0}&\cellcolor[gray]{0}&\cellcolor[gray]{0}&\cellcolor[gray]{0}\\
\hline
$-$&\cellcolor{Yellow}&10&7&10\\
\hline
$\pm$&\cellcolor[gray]{0}&\cellcolor[gray]{0}&\cellcolor[gray]{0}&\cellcolor[gray]{0}\\
\hline
\end{tabular}\renewcommand{\arraystretch}{1.5}
&\renewcommand{\arraystretch}{1.5}
\begin{tabular}{|M|M|M|M|M|}
\hline
$\rightarrow$ &$0$&$+$&$-$&$\pm$\\
\hline
$0$&\cellcolor[gray]{0}&\cellcolor[gray]{0}&\cellcolor[gray]{0}&\cellcolor[gray]{0}\\
\hline
$+$&\cellcolor[gray]{0}&\cellcolor[gray]{0}&\cellcolor[gray]{0}&\cellcolor[gray]{0}\\
\hline
$-$&7&9&10&4\\
\hline
$\pm$&\cellcolor[gray]{0}&\cellcolor[gray]{0}&\cellcolor[gray]{0}&\cellcolor[gray]{0}\\
\hline
\end{tabular}\renewcommand{\arraystretch}{1.5}
&\renewcommand{\arraystretch}{1.5}
\begin{tabular}{|M|M|M|M|M|}
\hline
$\rightarrow$ &$0$&$+$&$-$&$\pm$\\
\hline
$0$&\cellcolor[gray]{0}&\cellcolor[gray]{0}&\cellcolor[gray]{0}&\cellcolor[gray]{0}\\
\hline
$+$&\cellcolor[gray]{0}&\cellcolor[gray]{0}&\cellcolor[gray]{0}&\cellcolor[gray]{0}\\
\hline
$-$&10&8&10&\cellcolor{Orange}\\
\hline
$\pm$&\cellcolor[gray]{0}&\cellcolor[gray]{0}&\cellcolor[gray]{0}&\cellcolor[gray]{0}\\
\hline
\end{tabular}\renewcommand{\arraystretch}{1.5}
&\renewcommand{\arraystretch}{1.5}
\begin{tabular}{|M|M|M|M|M|}
\hline
$\rightarrow$ &$0$&$+$&$-$&$\pm$\\
\hline
$0$&\cellcolor[gray]{0}&\cellcolor[gray]{0}&\cellcolor[gray]{0}&\cellcolor[gray]{0}\\
\hline
$+$&\cellcolor[gray]{0}&\cellcolor[gray]{0}&\cellcolor[gray]{0}&\cellcolor[gray]{0}\\
\hline
$-$&10&8&10&\cellcolor{Orange}\\
\hline
$\pm$&\cellcolor[gray]{0}&\cellcolor[gray]{0}&\cellcolor[gray]{0}&\cellcolor[gray]{0}\\
\hline
\end{tabular}\renewcommand{\arraystretch}{1.5}
&\renewcommand{\arraystretch}{1.5}
\begin{tabular}{|M|M|M|M|M|}
\hline
$\rightarrow$ &$0$&$+$&$-$&$\pm$\\
\hline
$0$&\cellcolor[gray]{0}&\cellcolor[gray]{0}&\cellcolor[gray]{0}&\cellcolor[gray]{0}\\
\hline
$+$&\cellcolor[gray]{0}&\cellcolor[gray]{0}&\cellcolor[gray]{0}&\cellcolor[gray]{0}\\
\hline
$-$&10&4&10&6\\
\hline
$\pm$&\cellcolor[gray]{0}&\cellcolor[gray]{0}&\cellcolor[gray]{0}&\cellcolor[gray]{0}\\
\hline
\end{tabular}\renewcommand{\arraystretch}{1.5}
&\renewcommand{\arraystretch}{1.5}
\begin{tabular}{|M|M|M|M|M|}
\hline
$\rightarrow$ &$0$&$+$&$-$&$\pm$\\
\hline
$0$&\cellcolor[gray]{0}&\cellcolor[gray]{0}&\cellcolor[gray]{0}&\cellcolor[gray]{0}\\
\hline
$+$&\cellcolor[gray]{0}&\cellcolor[gray]{0}&\cellcolor[gray]{0}&\cellcolor[gray]{0}\\
\hline
$-$&\cellcolor{Yellow}&7&10&10\\
\hline
$\pm$&\cellcolor[gray]{0}&\cellcolor[gray]{0}&\cellcolor[gray]{0}&\cellcolor[gray]{0}\\
\hline
\end{tabular}\renewcommand{\arraystretch}{1.5}
&\renewcommand{\arraystretch}{1.5}
\begin{tabular}{|M|M|M|M|M|}
\hline
$\rightarrow$ &$0$&$+$&$-$&$\pm$\\
\hline
$0$&\cellcolor[gray]{0}&\cellcolor[gray]{0}&\cellcolor[gray]{0}&\cellcolor[gray]{0}\\
\hline
$+$&\cellcolor[gray]{0}&\cellcolor[gray]{0}&\cellcolor[gray]{0}&\cellcolor[gray]{0}\\
\hline
$-$&10&\cellcolor{Yellow}&10&7\\
\hline
$\pm$&\cellcolor[gray]{0}&\cellcolor[gray]{0}&\cellcolor[gray]{0}&\cellcolor[gray]{0}\\
\hline
\end{tabular}\renewcommand{\arraystretch}{1.5}
\\
&&&&&&&&\\
{\Large s}&\renewcommand{\arraystretch}{1.5}
\begin{tabular}{|M|M|M|M|M|}
\hline
$\rightarrow$ &$0$&$+$&$-$&$\pm$\\
\hline
$0$&7&7&\cellcolor[gray]{0}&7\\
\hline
$+$&\cellcolor[gray]{0}&\cellcolor[gray]{0}&\cellcolor[gray]{0}&\cellcolor[gray]{0}\\
\hline
$-$&\cellcolor{Yellow}&\cellcolor{Yellow}&\cellcolor[gray]{0}&\cellcolor{Yellow}\\
\hline
$\pm$&\cellcolor[gray]{0}&\cellcolor[gray]{0}&\cellcolor[gray]{0}&\cellcolor[gray]{0}\\
\hline
\end{tabular}\renewcommand{\arraystretch}{1.5}
&\renewcommand{\arraystretch}{1.5}
\begin{tabular}{|M|M|M|M|M|}
\hline
$\rightarrow$ &$0$&$+$&$-$&$\pm$\\
\hline
$0$&\cellcolor[gray]{0}&7&7&7\\
\hline
$+$&\cellcolor[gray]{0}&\cellcolor[gray]{0}&\cellcolor[gray]{0}&\cellcolor[gray]{0}\\
\hline
$-$&\cellcolor{Yellow}&\cellcolor{Yellow}&\cellcolor[gray]{0}&\cellcolor{Yellow}\\
\hline
$\pm$&\cellcolor[gray]{0}&\cellcolor[gray]{0}&\cellcolor[gray]{0}&\cellcolor[gray]{0}\\
\hline
\end{tabular}\renewcommand{\arraystretch}{1.5}
&\renewcommand{\arraystretch}{1.5}
\begin{tabular}{|M|M|M|M|M|}
\hline
$\rightarrow$ &$0$&$+$&$-$&$\pm$\\
\hline
$0$&5&6&7&\cellcolor{Yellow}\\
\hline
$+$&\cellcolor[gray]{0}&\cellcolor[gray]{0}&\cellcolor[gray]{0}&\cellcolor[gray]{0}\\
\hline
$-$&\cellcolor{Orange}&\cellcolor{Yellow}&\cellcolor{Yellow}&\cellcolor{Red}\\
\hline
$\pm$&\cellcolor[gray]{0}&\cellcolor[gray]{0}&\cellcolor[gray]{0}&\cellcolor[gray]{0}\\
\hline
\end{tabular}\renewcommand{\arraystretch}{1.5}
&\renewcommand{\arraystretch}{1.5}
\begin{tabular}{|M|M|M|M|M|}
\hline
$\rightarrow$ &$0$&$+$&$-$&$\pm$\\
\hline
$0$&7&6&7&\cellcolor{Red}\\
\hline
$+$&\cellcolor[gray]{0}&\cellcolor[gray]{0}&\cellcolor[gray]{0}&\cellcolor[gray]{0}\\
\hline
$-$&\cellcolor{Yellow}&\cellcolor{Orange}&\cellcolor{Yellow}&\cellcolor{Red}\\
\hline
$\pm$&\cellcolor[gray]{0}&\cellcolor[gray]{0}&\cellcolor[gray]{0}&\cellcolor[gray]{0}\\
\hline
\end{tabular}\renewcommand{\arraystretch}{1.5}
&\renewcommand{\arraystretch}{1.5}
\begin{tabular}{|M|M|M|M|M|}
\hline
$\rightarrow$ &$0$&$+$&$-$&$\pm$\\
\hline
$0$&7&7&7&\cellcolor[gray]{0}\\
\hline
$+$&\cellcolor[gray]{0}&\cellcolor[gray]{0}&\cellcolor[gray]{0}&\cellcolor[gray]{0}\\
\hline
$-$&\cellcolor{Yellow}&\cellcolor{Red}&\cellcolor{Yellow}&\cellcolor{Orange}\\
\hline
$\pm$&\cellcolor[gray]{0}&\cellcolor[gray]{0}&\cellcolor[gray]{0}&\cellcolor[gray]{0}\\
\hline
\end{tabular}\renewcommand{\arraystretch}{1.5}
&\renewcommand{\arraystretch}{1.5}
\begin{tabular}{|M|M|M|M|M|}
\hline
$\rightarrow$ &$0$&$+$&$-$&$\pm$\\
\hline
$0$&7&4&7&\cellcolor{Yellow}\\
\hline
$+$&\cellcolor[gray]{0}&\cellcolor[gray]{0}&\cellcolor[gray]{0}&\cellcolor[gray]{0}\\
\hline
$-$&\cellcolor{Yellow}&\cellcolor[gray]{0}&\cellcolor{Yellow}&\cellcolor{Yellow}\\
\hline
$\pm$&\cellcolor[gray]{0}&\cellcolor[gray]{0}&\cellcolor[gray]{0}&\cellcolor[gray]{0}\\
\hline
\end{tabular}\renewcommand{\arraystretch}{1.5}
&\renewcommand{\arraystretch}{1.5}
\begin{tabular}{|M|M|M|M|M|}
\hline
$\rightarrow$ &$0$&$+$&$-$&$\pm$\\
\hline
$0$&\cellcolor{Red}&6&7&7\\
\hline
$+$&\cellcolor[gray]{0}&\cellcolor[gray]{0}&\cellcolor[gray]{0}&\cellcolor[gray]{0}\\
\hline
$-$&\cellcolor{Orange}&\cellcolor{Red}&\cellcolor{Yellow}&\cellcolor{Yellow}\\
\hline
$\pm$&\cellcolor[gray]{0}&\cellcolor[gray]{0}&\cellcolor[gray]{0}&\cellcolor[gray]{0}\\
\hline
\end{tabular}\renewcommand{\arraystretch}{1.5}
&\renewcommand{\arraystretch}{1.5}
\begin{tabular}{|M|M|M|M|M|}
\hline
$\rightarrow$ &$0$&$+$&$-$&$\pm$\\
\hline
$0$&7&\cellcolor{Red}&7&6\\
\hline
$+$&\cellcolor[gray]{0}&\cellcolor[gray]{0}&\cellcolor[gray]{0}&\cellcolor[gray]{0}\\
\hline
$-$&\cellcolor{Yellow}&\cellcolor{Orange}&\cellcolor{Yellow}&\cellcolor{Red}\\
\hline
$\pm$&\cellcolor[gray]{0}&\cellcolor[gray]{0}&\cellcolor[gray]{0}&\cellcolor[gray]{0}\\
\hline
\end{tabular}\renewcommand{\arraystretch}{1.5}
\\
&&&&&&&&\\[-2\jot]
\hline
&&&&&&&&\\[-2\jot]
{\Large e}&\renewcommand{\arraystretch}{1.5}
\begin{tabular}{|M|M|M|M|M|}
\hline
$\rightarrow$ &$0$&$+$&$-$&$\pm$\\
\hline
$0$&\cellcolor{Yellow}&\cellcolor{Yellow}&\cellcolor[gray]{0}&\cellcolor{Yellow}\\
\hline
$+$&\cellcolor{Red}&\cellcolor{Red}&\cellcolor[gray]{0}&\cellcolor{Red}\\
\hline
$-$&\cellcolor[gray]{0}&\cellcolor[gray]{0}&\cellcolor[gray]{0}&\cellcolor[gray]{0}\\
\hline
$\pm$&6&6&\cellcolor[gray]{0}&6\\
\hline
\end{tabular}\renewcommand{\arraystretch}{1.5}
&\renewcommand{\arraystretch}{1.5}
\begin{tabular}{|M|M|M|M|M|}
\hline
$\rightarrow$ &$0$&$+$&$-$&$\pm$\\
\hline
$0$&\cellcolor{Red}&\cellcolor{Yellow}&\cellcolor{Orange}&\cellcolor{Yellow}\\
\hline
$+$&\cellcolor[gray]{0}&\cellcolor{Red}&\cellcolor{Red}&\cellcolor{Red}\\
\hline
$-$&\cellcolor[gray]{0}&\cellcolor[gray]{0}&\cellcolor[gray]{0}&\cellcolor[gray]{0}\\
\hline
$\pm$&\cellcolor{Orange}&6&4&6\\
\hline
\end{tabular}\renewcommand{\arraystretch}{1.5}
&\renewcommand{\arraystretch}{1.5}
\begin{tabular}{|M|M|M|M|M|}
\hline
$\rightarrow$ &$0$&$+$&$-$&$\pm$\\
\hline
$0$&\cellcolor[gray]{0}&\cellcolor{Yellow}&\cellcolor{Yellow}&\cellcolor{Yellow}\\
\hline
$+$&\cellcolor{Red}&\cellcolor[gray]{0}&\cellcolor{Red}&\cellcolor{Red}\\
\hline
$-$&\cellcolor[gray]{0}&\cellcolor[gray]{0}&\cellcolor[gray]{0}&\cellcolor[gray]{0}\\
\hline
$\pm$&6&6&6&\cellcolor[gray]{0}\\
\hline
\end{tabular}\renewcommand{\arraystretch}{1.5}
&\renewcommand{\arraystretch}{1.5}
\begin{tabular}{|M|M|M|M|M|}
\hline
$\rightarrow$ &$0$&$+$&$-$&$\pm$\\
\hline
$0$&\cellcolor{Yellow}&\cellcolor{Orange}&\cellcolor{Yellow}&\cellcolor{Red}\\
\hline
$+$&\cellcolor{Red}&\cellcolor{Red}&\cellcolor{Red}&\cellcolor[gray]{0}\\
\hline
$-$&\cellcolor[gray]{0}&\cellcolor[gray]{0}&\cellcolor[gray]{0}&\cellcolor[gray]{0}\\
\hline
$\pm$&6&5&6&\cellcolor{Red}\\
\hline
\end{tabular}\renewcommand{\arraystretch}{1.5}
&\renewcommand{\arraystretch}{1.5}
\begin{tabular}{|M|M|M|M|M|}
\hline
$\rightarrow$ &$0$&$+$&$-$&$\pm$\\
\hline
$0$&\cellcolor{Yellow}&\cellcolor{Orange}&\cellcolor{Yellow}&\cellcolor{Red}\\
\hline
$+$&\cellcolor{Red}&\cellcolor{Red}&\cellcolor{Red}&\cellcolor[gray]{0}\\
\hline
$-$&\cellcolor[gray]{0}&\cellcolor[gray]{0}&\cellcolor[gray]{0}&\cellcolor[gray]{0}\\
\hline
$\pm$&6&5&6&\cellcolor{Red}\\
\hline
\end{tabular}\renewcommand{\arraystretch}{1.5}
&\renewcommand{\arraystretch}{1.5}
\begin{tabular}{|M|M|M|M|M|}
\hline
$\rightarrow$ &$0$&$+$&$-$&$\pm$\\
\hline
$0$&\cellcolor{Yellow}&\cellcolor{Red}&\cellcolor{Yellow}&\cellcolor{Orange}\\
\hline
$+$&\cellcolor{Red}&\cellcolor[gray]{0}&\cellcolor{Red}&\cellcolor{Red}\\
\hline
$-$&\cellcolor[gray]{0}&\cellcolor[gray]{0}&\cellcolor[gray]{0}&\cellcolor[gray]{0}\\
\hline
$\pm$&6&\cellcolor{Yellow}&6&\cellcolor{Yellow}\\
\hline
\end{tabular}\renewcommand{\arraystretch}{1.5}
&\renewcommand{\arraystretch}{1.5}
\begin{tabular}{|M|M|M|M|M|}
\hline
$\rightarrow$ &$0$&$+$&$-$&$\pm$\\
\hline
$0$&\cellcolor{Red}&\cellcolor{Orange}&\cellcolor{Yellow}&\cellcolor{Yellow}\\
\hline
$+$&\cellcolor[gray]{0}&\cellcolor{Red}&\cellcolor{Red}&\cellcolor{Red}\\
\hline
$-$&\cellcolor[gray]{0}&\cellcolor[gray]{0}&\cellcolor[gray]{0}&\cellcolor[gray]{0}\\
\hline
$\pm$&\cellcolor{Orange}&4&6&6\\
\hline
\end{tabular}\renewcommand{\arraystretch}{1.5}
&\renewcommand{\arraystretch}{1.5}
\begin{tabular}{|M|M|M|M|M|}
\hline
$\rightarrow$ &$0$&$+$&$-$&$\pm$\\
\hline
$0$&\cellcolor{Yellow}&\cellcolor{Red}&\cellcolor{Yellow}&\cellcolor{Orange}\\
\hline
$+$&\cellcolor{Red}&\cellcolor[gray]{0}&\cellcolor{Red}&\cellcolor{Red}\\
\hline
$-$&\cellcolor[gray]{0}&\cellcolor[gray]{0}&\cellcolor[gray]{0}&\cellcolor[gray]{0}\\
\hline
$\pm$&6&\cellcolor{Orange}&6&4\\
\hline
\end{tabular}\renewcommand{\arraystretch}{1.5}
\\
&&&&&&&&\\
{\Large hy}&\renewcommand{\arraystretch}{1.5}
\begin{tabular}{|M|M|M|M|M|}
\hline
$\rightarrow$ &$0$&$+$&$-$&$\pm$\\
\hline
$0$&\cellcolor[gray]{0}&\cellcolor[gray]{0}&\cellcolor[gray]{0}&\cellcolor[gray]{0}\\
\hline
$+$&\cellcolor{Orange}&\cellcolor{Orange}&\cellcolor[gray]{0}&\cellcolor{Orange}\\
\hline
$-$&\cellcolor[gray]{0}&\cellcolor[gray]{0}&\cellcolor[gray]{0}&\cellcolor[gray]{0}\\
\hline
$\pm$&8&8&\cellcolor[gray]{0}&8\\
\hline
\end{tabular}\renewcommand{\arraystretch}{1.5}
&\renewcommand{\arraystretch}{1.5}
\begin{tabular}{|M|M|M|M|M|}
\hline
$\rightarrow$ &$0$&$+$&$-$&$\pm$\\
\hline
$0$&\cellcolor[gray]{0}&\cellcolor[gray]{0}&\cellcolor[gray]{0}&\cellcolor[gray]{0}\\
\hline
$+$&\cellcolor{Red}&\cellcolor{Orange}&\cellcolor{Red}&\cellcolor{Orange}\\
\hline
$-$&\cellcolor[gray]{0}&\cellcolor[gray]{0}&\cellcolor[gray]{0}&\cellcolor[gray]{0}\\
\hline
$\pm$&\cellcolor{Orange}&8&6&8\\
\hline
\end{tabular}\renewcommand{\arraystretch}{1.5}
&\renewcommand{\arraystretch}{1.5}
\begin{tabular}{|M|M|M|M|M|}
\hline
$\rightarrow$ &$0$&$+$&$-$&$\pm$\\
\hline
$0$&\cellcolor[gray]{0}&\cellcolor[gray]{0}&\cellcolor[gray]{0}&\cellcolor[gray]{0}\\
\hline
$+$&\cellcolor{Red}&\cellcolor{Orange}&\cellcolor{Orange}&\cellcolor{Red}\\
\hline
$-$&\cellcolor[gray]{0}&\cellcolor[gray]{0}&\cellcolor[gray]{0}&\cellcolor[gray]{0}\\
\hline
$\pm$&6&7&8&\cellcolor{Yellow}\\
\hline
\end{tabular}\renewcommand{\arraystretch}{1.5}
&\renewcommand{\arraystretch}{1.5}
\begin{tabular}{|M|M|M|M|M|}
\hline
$\rightarrow$ &$0$&$+$&$-$&$\pm$\\
\hline
$0$&\cellcolor[gray]{0}&\cellcolor[gray]{0}&\cellcolor[gray]{0}&\cellcolor[gray]{0}\\
\hline
$+$&\cellcolor{Orange}&\cellcolor[gray]{0}&\cellcolor{Orange}&\cellcolor{Orange}\\
\hline
$-$&\cellcolor[gray]{0}&\cellcolor[gray]{0}&\cellcolor[gray]{0}&\cellcolor[gray]{0}\\
\hline
$\pm$&8&8&8&\cellcolor[gray]{0}\\
\hline
\end{tabular}\renewcommand{\arraystretch}{1.5}
&\renewcommand{\arraystretch}{1.5}
\begin{tabular}{|M|M|M|M|M|}
\hline
$\rightarrow$ &$0$&$+$&$-$&$\pm$\\
\hline
$0$&\cellcolor[gray]{0}&\cellcolor[gray]{0}&\cellcolor[gray]{0}&\cellcolor[gray]{0}\\
\hline
$+$&\cellcolor{Orange}&\cellcolor{Red}&\cellcolor{Orange}&\cellcolor{Red}\\
\hline
$-$&\cellcolor[gray]{0}&\cellcolor[gray]{0}&\cellcolor[gray]{0}&\cellcolor[gray]{0}\\
\hline
$\pm$&8&7&8&\cellcolor{Red}\\
\hline
\end{tabular}\renewcommand{\arraystretch}{1.5}
&\renewcommand{\arraystretch}{1.5}
\begin{tabular}{|M|M|M|M|M|}
\hline
$\rightarrow$ &$0$&$+$&$-$&$\pm$\\
\hline
$0$&\cellcolor[gray]{0}&\cellcolor[gray]{0}&\cellcolor[gray]{0}&\cellcolor[gray]{0}\\
\hline
$+$&\cellcolor{Orange}&\cellcolor{Red}&\cellcolor{Orange}&\cellcolor{Red}\\
\hline
$-$&\cellcolor[gray]{0}&\cellcolor[gray]{0}&\cellcolor[gray]{0}&\cellcolor[gray]{0}\\
\hline
$\pm$&8&\cellcolor{Yellow}&8&5\\
\hline
\end{tabular}\renewcommand{\arraystretch}{1.5}
&\renewcommand{\arraystretch}{1.5}
\begin{tabular}{|M|M|M|M|M|}
\hline
$\rightarrow$ &$0$&$+$&$-$&$\pm$\\
\hline
$0$&\cellcolor[gray]{0}&\cellcolor[gray]{0}&\cellcolor[gray]{0}&\cellcolor[gray]{0}\\
\hline
$+$&\cellcolor[gray]{0}&\cellcolor{Orange}&\cellcolor{Orange}&\cellcolor{Orange}\\
\hline
$-$&\cellcolor[gray]{0}&\cellcolor[gray]{0}&\cellcolor[gray]{0}&\cellcolor[gray]{0}\\
\hline
$\pm$&\cellcolor{Yellow}&5&8&8\\
\hline
\end{tabular}\renewcommand{\arraystretch}{1.5}
&\renewcommand{\arraystretch}{1.5}
\begin{tabular}{|M|M|M|M|M|}
\hline
$\rightarrow$ &$0$&$+$&$-$&$\pm$\\
\hline
$0$&\cellcolor[gray]{0}&\cellcolor[gray]{0}&\cellcolor[gray]{0}&\cellcolor[gray]{0}\\
\hline
$+$&\cellcolor{Orange}&\cellcolor[gray]{0}&\cellcolor{Orange}&\cellcolor{Orange}\\
\hline
$-$&\cellcolor[gray]{0}&\cellcolor[gray]{0}&\cellcolor[gray]{0}&\cellcolor[gray]{0}\\
\hline
$\pm$&8&\cellcolor{Yellow}&8&5\\
\hline
\end{tabular}\renewcommand{\arraystretch}{1.5}
\\
&&&&&&&&\\
{\Large k}&\renewcommand{\arraystretch}{1.5}
\begin{tabular}{|M|M|M|M|M|}
\hline
$\rightarrow$ &$0$&$+$&$-$&$\pm$\\
\hline
$0$&\cellcolor[gray]{0}&\cellcolor[gray]{0}&\cellcolor[gray]{0}&\cellcolor[gray]{0}\\
\hline
$+$&\cellcolor{Orange}&\cellcolor{Orange}&\cellcolor[gray]{0}&\cellcolor{Orange}\\
\hline
$-$&\cellcolor[gray]{0}&\cellcolor[gray]{0}&\cellcolor[gray]{0}&\cellcolor[gray]{0}\\
\hline
$\pm$&8&8&\cellcolor[gray]{0}&8\\
\hline
\end{tabular}\renewcommand{\arraystretch}{1.5}
&\renewcommand{\arraystretch}{1.5}
\begin{tabular}{|M|M|M|M|M|}
\hline
$\rightarrow$ &$0$&$+$&$-$&$\pm$\\
\hline
$0$&\cellcolor[gray]{0}&\cellcolor[gray]{0}&\cellcolor[gray]{0}&\cellcolor[gray]{0}\\
\hline
$+$&\cellcolor{Orange}&\cellcolor{Orange}&\cellcolor[gray]{0}&\cellcolor{Orange}\\
\hline
$-$&\cellcolor[gray]{0}&\cellcolor[gray]{0}&\cellcolor[gray]{0}&\cellcolor[gray]{0}\\
\hline
$\pm$&\cellcolor{Red}&8&7&8\\
\hline
\end{tabular}\renewcommand{\arraystretch}{1.5}
&\renewcommand{\arraystretch}{1.5}
\begin{tabular}{|M|M|M|M|M|}
\hline
$\rightarrow$ &$0$&$+$&$-$&$\pm$\\
\hline
$0$&\cellcolor[gray]{0}&\cellcolor[gray]{0}&\cellcolor[gray]{0}&\cellcolor[gray]{0}\\
\hline
$+$&\cellcolor{Red}&\cellcolor{Orange}&\cellcolor{Orange}&\cellcolor{Red}\\
\hline
$-$&\cellcolor[gray]{0}&\cellcolor[gray]{0}&\cellcolor[gray]{0}&\cellcolor[gray]{0}\\
\hline
$\pm$&6&7&8&\cellcolor{Yellow}\\
\hline
\end{tabular}\renewcommand{\arraystretch}{1.5}
&\renewcommand{\arraystretch}{1.5}
\begin{tabular}{|M|M|M|M|M|}
\hline
$\rightarrow$ &$0$&$+$&$-$&$\pm$\\
\hline
$0$&\cellcolor[gray]{0}&\cellcolor[gray]{0}&\cellcolor[gray]{0}&\cellcolor[gray]{0}\\
\hline
$+$&\cellcolor{Orange}&\cellcolor{Red}&\cellcolor{Orange}&\cellcolor{Red}\\
\hline
$-$&\cellcolor[gray]{0}&\cellcolor[gray]{0}&\cellcolor[gray]{0}&\cellcolor[gray]{0}\\
\hline
$\pm$&8&7&8&\cellcolor{Red}\\
\hline
\end{tabular}\renewcommand{\arraystretch}{1.5}
&\renewcommand{\arraystretch}{1.5}
\begin{tabular}{|M|M|M|M|M|}
\hline
$\rightarrow$ &$0$&$+$&$-$&$\pm$\\
\hline
$0$&\cellcolor[gray]{0}&\cellcolor[gray]{0}&\cellcolor[gray]{0}&\cellcolor[gray]{0}\\
\hline
$+$&\cellcolor{Orange}&\cellcolor[gray]{0}&\cellcolor{Orange}&\cellcolor{Orange}\\
\hline
$-$&\cellcolor[gray]{0}&\cellcolor[gray]{0}&\cellcolor[gray]{0}&\cellcolor[gray]{0}\\
\hline
$\pm$&8&8&8&\cellcolor[gray]{0}\\
\hline
\end{tabular}\renewcommand{\arraystretch}{1.5}
&\renewcommand{\arraystretch}{1.5}
\begin{tabular}{|M|M|M|M|M|}
\hline
$\rightarrow$ &$0$&$+$&$-$&$\pm$\\
\hline
$0$&\cellcolor[gray]{0}&\cellcolor[gray]{0}&\cellcolor[gray]{0}&\cellcolor[gray]{0}\\
\hline
$+$&\cellcolor{Orange}&\cellcolor[gray]{0}&\cellcolor{Orange}&\cellcolor{Orange}\\
\hline
$-$&\cellcolor[gray]{0}&\cellcolor[gray]{0}&\cellcolor[gray]{0}&\cellcolor[gray]{0}\\
\hline
$\pm$&8&4&8&4\\
\hline
\end{tabular}\renewcommand{\arraystretch}{1.5}
&\renewcommand{\arraystretch}{1.5}
\begin{tabular}{|M|M|M|M|M|}
\hline
$\rightarrow$ &$0$&$+$&$-$&$\pm$\\
\hline
$0$&\cellcolor[gray]{0}&\cellcolor[gray]{0}&\cellcolor[gray]{0}&\cellcolor[gray]{0}\\
\hline
$+$&\cellcolor{Red}&\cellcolor{Red}&\cellcolor{Orange}&\cellcolor{Orange}\\
\hline
$-$&\cellcolor[gray]{0}&\cellcolor[gray]{0}&\cellcolor[gray]{0}&\cellcolor[gray]{0}\\
\hline
$\pm$&\cellcolor{Orange}&6&8&8\\
\hline
\end{tabular}\renewcommand{\arraystretch}{1.5}
&\renewcommand{\arraystretch}{1.5}
\begin{tabular}{|M|M|M|M|M|}
\hline
$\rightarrow$ &$0$&$+$&$-$&$\pm$\\
\hline
$0$&\cellcolor[gray]{0}&\cellcolor[gray]{0}&\cellcolor[gray]{0}&\cellcolor[gray]{0}\\
\hline
$+$&\cellcolor{Orange}&\cellcolor{Red}&\cellcolor{Orange}&\cellcolor{Red}\\
\hline
$-$&\cellcolor[gray]{0}&\cellcolor[gray]{0}&\cellcolor[gray]{0}&\cellcolor[gray]{0}\\
\hline
$\pm$&8&\cellcolor{Orange}&8&6\\
\hline
\end{tabular}\renewcommand{\arraystretch}{1.5}
\\
&&&&&&&&\\
{\Large p}&\renewcommand{\arraystretch}{1.5}
\begin{tabular}{|M|M|M|M|M|}
\hline
$\rightarrow$ &$0$&$+$&$-$&$\pm$\\
\hline
$0$&\cellcolor[gray]{0}&\cellcolor[gray]{0}&\cellcolor[gray]{0}&\cellcolor[gray]{0}\\
\hline
$+$&6&6&\cellcolor[gray]{0}&6\\
\hline
$-$&\cellcolor[gray]{0}&\cellcolor[gray]{0}&\cellcolor[gray]{0}&\cellcolor[gray]{0}\\
\hline
$\pm$&4&4&\cellcolor[gray]{0}&4\\
\hline
\end{tabular}\renewcommand{\arraystretch}{1.5}
&\renewcommand{\arraystretch}{1.5}
\begin{tabular}{|M|M|M|M|M|}
\hline
$\rightarrow$ &$0$&$+$&$-$&$\pm$\\
\hline
$0$&\cellcolor[gray]{0}&\cellcolor[gray]{0}&\cellcolor[gray]{0}&\cellcolor[gray]{0}\\
\hline
$+$&\cellcolor{Yellow}&6&\cellcolor{Yellow}&6\\
\hline
$-$&\cellcolor[gray]{0}&\cellcolor[gray]{0}&\cellcolor[gray]{0}&\cellcolor[gray]{0}\\
\hline
$\pm$&\cellcolor[gray]{0}&4&4&4\\
\hline
\end{tabular}\renewcommand{\arraystretch}{1.5}
&\renewcommand{\arraystretch}{1.5}
\begin{tabular}{|M|M|M|M|M|}
\hline
$\rightarrow$ &$0$&$+$&$-$&$\pm$\\
\hline
$0$&\cellcolor[gray]{0}&\cellcolor[gray]{0}&\cellcolor[gray]{0}&\cellcolor[gray]{0}\\
\hline
$+$&4&5&6&\cellcolor{Yellow}\\
\hline
$-$&\cellcolor[gray]{0}&\cellcolor[gray]{0}&\cellcolor[gray]{0}&\cellcolor[gray]{0}\\
\hline
$\pm$&\cellcolor{Yellow}&4&4&\cellcolor{Red}\\
\hline
\end{tabular}\renewcommand{\arraystretch}{1.5}
&\renewcommand{\arraystretch}{1.5}
\begin{tabular}{|M|M|M|M|M|}
\hline
$\rightarrow$ &$0$&$+$&$-$&$\pm$\\
\hline
$0$&\cellcolor[gray]{0}&\cellcolor[gray]{0}&\cellcolor[gray]{0}&\cellcolor[gray]{0}\\
\hline
$+$&6&5&6&\cellcolor{Red}\\
\hline
$-$&\cellcolor[gray]{0}&\cellcolor[gray]{0}&\cellcolor[gray]{0}&\cellcolor[gray]{0}\\
\hline
$\pm$&4&\cellcolor{Yellow}&4&\cellcolor{Red}\\
\hline
\end{tabular}\renewcommand{\arraystretch}{1.5}
&\renewcommand{\arraystretch}{1.5}
\begin{tabular}{|M|M|M|M|M|}
\hline
$\rightarrow$ &$0$&$+$&$-$&$\pm$\\
\hline
$0$&\cellcolor[gray]{0}&\cellcolor[gray]{0}&\cellcolor[gray]{0}&\cellcolor[gray]{0}\\
\hline
$+$&6&4&6&\cellcolor{Orange}\\
\hline
$-$&\cellcolor[gray]{0}&\cellcolor[gray]{0}&\cellcolor[gray]{0}&\cellcolor[gray]{0}\\
\hline
$\pm$&4&4&4&\cellcolor[gray]{0}\\
\hline
\end{tabular}\renewcommand{\arraystretch}{1.5}
&\renewcommand{\arraystretch}{1.5}
\begin{tabular}{|M|M|M|M|M|}
\hline
$\rightarrow$ &$0$&$+$&$-$&$\pm$\\
\hline
$0$&\cellcolor[gray]{0}&\cellcolor[gray]{0}&\cellcolor[gray]{0}&\cellcolor[gray]{0}\\
\hline
$+$&6&\cellcolor[gray]{0}&6&6\\
\hline
$-$&\cellcolor[gray]{0}&\cellcolor[gray]{0}&\cellcolor[gray]{0}&\cellcolor[gray]{0}\\
\hline
$\pm$&4&4&4&\cellcolor[gray]{0}\\
\hline
\end{tabular}\renewcommand{\arraystretch}{1.5}
&\renewcommand{\arraystretch}{1.5}
\begin{tabular}{|M|M|M|M|M|}
\hline
$\rightarrow$ &$0$&$+$&$-$&$\pm$\\
\hline
$0$&\cellcolor[gray]{0}&\cellcolor[gray]{0}&\cellcolor[gray]{0}&\cellcolor[gray]{0}\\
\hline
$+$&\cellcolor{Orange}&4&6&6\\
\hline
$-$&\cellcolor[gray]{0}&\cellcolor[gray]{0}&\cellcolor[gray]{0}&\cellcolor[gray]{0}\\
\hline
$\pm$&\cellcolor{Red}&\cellcolor{Yellow}&4&4\\
\hline
\end{tabular}\renewcommand{\arraystretch}{1.5}
&\renewcommand{\arraystretch}{1.5}
\begin{tabular}{|M|M|M|M|M|}
\hline
$\rightarrow$ &$0$&$+$&$-$&$\pm$\\
\hline
$0$&\cellcolor[gray]{0}&\cellcolor[gray]{0}&\cellcolor[gray]{0}&\cellcolor[gray]{0}\\
\hline
$+$&6&\cellcolor{Orange}&6&4\\
\hline
$-$&\cellcolor[gray]{0}&\cellcolor[gray]{0}&\cellcolor[gray]{0}&\cellcolor[gray]{0}\\
\hline
$\pm$&4&\cellcolor{Red}&4&\cellcolor{Yellow}\\
\hline
\end{tabular}\renewcommand{\arraystretch}{1.5}
\\
&&&&&&&&\\[-2\jot]
\hline
&&&&&&&&\\[-2\jot]
{\Large d}&\renewcommand{\arraystretch}{1.5}
\begin{tabular}{|M|M|M|M|M|}
\hline
$\rightarrow$ &$0$&$+$&$-$&$\pm$\\
\hline
$0$&7&7&\cellcolor[gray]{0}&7\\
\hline
$+$&\cellcolor{Yellow}&\cellcolor{Yellow}&\cellcolor[gray]{0}&\cellcolor{Yellow}\\
\hline
$-$&\cellcolor[gray]{0}&\cellcolor[gray]{0}&\cellcolor[gray]{0}&\cellcolor[gray]{0}\\
\hline
$\pm$&\cellcolor[gray]{0}&\cellcolor[gray]{0}&\cellcolor[gray]{0}&\cellcolor[gray]{0}\\
\hline
\end{tabular}\renewcommand{\arraystretch}{1.5}
&\renewcommand{\arraystretch}{1.5}
\begin{tabular}{|M|M|M|M|M|}
\hline
$\rightarrow$ &$0$&$+$&$-$&$\pm$\\
\hline
$0$&\cellcolor{Red}&7&6&7\\
\hline
$+$&\cellcolor{Orange}&\cellcolor{Yellow}&\cellcolor{Red}&\cellcolor{Yellow}\\
\hline
$-$&\cellcolor[gray]{0}&\cellcolor[gray]{0}&\cellcolor[gray]{0}&\cellcolor[gray]{0}\\
\hline
$\pm$&\cellcolor[gray]{0}&\cellcolor[gray]{0}&\cellcolor[gray]{0}&\cellcolor[gray]{0}\\
\hline
\end{tabular}\renewcommand{\arraystretch}{1.5}
&\renewcommand{\arraystretch}{1.5}
\begin{tabular}{|M|M|M|M|M|}
\hline
$\rightarrow$ &$0$&$+$&$-$&$\pm$\\
\hline
$0$&5&6&7&\cellcolor{Yellow}\\
\hline
$+$&\cellcolor{Orange}&\cellcolor{Yellow}&\cellcolor{Yellow}&\cellcolor{Red}\\
\hline
$-$&\cellcolor[gray]{0}&\cellcolor[gray]{0}&\cellcolor[gray]{0}&\cellcolor[gray]{0}\\
\hline
$\pm$&\cellcolor[gray]{0}&\cellcolor[gray]{0}&\cellcolor[gray]{0}&\cellcolor[gray]{0}\\
\hline
\end{tabular}\renewcommand{\arraystretch}{1.5}
&\renewcommand{\arraystretch}{1.5}
\begin{tabular}{|M|M|M|M|M|}
\hline
$\rightarrow$ &$0$&$+$&$-$&$\pm$\\
\hline
$0$&7&5&7&\cellcolor{Orange}\\
\hline
$+$&\cellcolor{Yellow}&\cellcolor{Yellow}&\cellcolor{Yellow}&\cellcolor[gray]{0}\\
\hline
$-$&\cellcolor[gray]{0}&\cellcolor[gray]{0}&\cellcolor[gray]{0}&\cellcolor[gray]{0}\\
\hline
$\pm$&\cellcolor[gray]{0}&\cellcolor[gray]{0}&\cellcolor[gray]{0}&\cellcolor[gray]{0}\\
\hline
\end{tabular}\renewcommand{\arraystretch}{1.5}
&\renewcommand{\arraystretch}{1.5}
\begin{tabular}{|M|M|M|M|M|}
\hline
$\rightarrow$ &$0$&$+$&$-$&$\pm$\\
\hline
$0$&7&6&7&\cellcolor{Red}\\
\hline
$+$&\cellcolor{Yellow}&\cellcolor{Orange}&\cellcolor{Yellow}&\cellcolor{Red}\\
\hline
$-$&\cellcolor[gray]{0}&\cellcolor[gray]{0}&\cellcolor[gray]{0}&\cellcolor[gray]{0}\\
\hline
$\pm$&\cellcolor[gray]{0}&\cellcolor[gray]{0}&\cellcolor[gray]{0}&\cellcolor[gray]{0}\\
\hline
\end{tabular}\renewcommand{\arraystretch}{1.5}
&\renewcommand{\arraystretch}{1.5}
\begin{tabular}{|M|M|M|M|M|}
\hline
$\rightarrow$ &$0$&$+$&$-$&$\pm$\\
\hline
$0$&7&\cellcolor{Yellow}&7&4\\
\hline
$+$&\cellcolor{Yellow}&\cellcolor{Red}&\cellcolor{Yellow}&\cellcolor{Orange}\\
\hline
$-$&\cellcolor[gray]{0}&\cellcolor[gray]{0}&\cellcolor[gray]{0}&\cellcolor[gray]{0}\\
\hline
$\pm$&\cellcolor[gray]{0}&\cellcolor[gray]{0}&\cellcolor[gray]{0}&\cellcolor[gray]{0}\\
\hline
\end{tabular}\renewcommand{\arraystretch}{1.5}
&\renewcommand{\arraystretch}{1.5}
\begin{tabular}{|M|M|M|M|M|}
\hline
$\rightarrow$ &$0$&$+$&$-$&$\pm$\\
\hline
$0$&\cellcolor[gray]{0}&7&7&7\\
\hline
$+$&\cellcolor{Yellow}&\cellcolor[gray]{0}&\cellcolor{Yellow}&\cellcolor{Yellow}\\
\hline
$-$&\cellcolor[gray]{0}&\cellcolor[gray]{0}&\cellcolor[gray]{0}&\cellcolor[gray]{0}\\
\hline
$\pm$&\cellcolor[gray]{0}&\cellcolor[gray]{0}&\cellcolor[gray]{0}&\cellcolor[gray]{0}\\
\hline
\end{tabular}\renewcommand{\arraystretch}{1.5}
&\renewcommand{\arraystretch}{1.5}
\begin{tabular}{|M|M|M|M|M|}
\hline
$\rightarrow$ &$0$&$+$&$-$&$\pm$\\
\hline
$0$&7&\cellcolor[gray]{0}&7&7\\
\hline
$+$&\cellcolor{Yellow}&\cellcolor{Yellow}&\cellcolor{Yellow}&\cellcolor[gray]{0}\\
\hline
$-$&\cellcolor[gray]{0}&\cellcolor[gray]{0}&\cellcolor[gray]{0}&\cellcolor[gray]{0}\\
\hline
$\pm$&\cellcolor[gray]{0}&\cellcolor[gray]{0}&\cellcolor[gray]{0}&\cellcolor[gray]{0}\\
\hline
\end{tabular}\renewcommand{\arraystretch}{1.5}
\\
&&&&&&&&\\
{\Large m}&\renewcommand{\arraystretch}{1.5}
\begin{tabular}{|M|M|M|M|M|}
\hline
$\rightarrow$ &$0$&$+$&$-$&$\pm$\\
\hline
$0$&\cellcolor[gray]{0}&\cellcolor[gray]{0}&\cellcolor[gray]{0}&\cellcolor[gray]{0}\\
\hline
$+$&7&7&\cellcolor[gray]{0}&7\\
\hline
$-$&\cellcolor[gray]{0}&\cellcolor[gray]{0}&\cellcolor[gray]{0}&\cellcolor[gray]{0}\\
\hline
$\pm$&\cellcolor{Yellow}&\cellcolor{Yellow}&\cellcolor[gray]{0}&\cellcolor{Yellow}\\
\hline
\end{tabular}\renewcommand{\arraystretch}{1.5}
&\renewcommand{\arraystretch}{1.5}
\begin{tabular}{|M|M|M|M|M|}
\hline
$\rightarrow$ &$0$&$+$&$-$&$\pm$\\
\hline
$0$&\cellcolor[gray]{0}&\cellcolor[gray]{0}&\cellcolor[gray]{0}&\cellcolor[gray]{0}\\
\hline
$+$&\cellcolor{Red}&7&6&7\\
\hline
$-$&\cellcolor[gray]{0}&\cellcolor[gray]{0}&\cellcolor[gray]{0}&\cellcolor[gray]{0}\\
\hline
$\pm$&\cellcolor{Orange}&\cellcolor{Yellow}&\cellcolor{Red}&\cellcolor{Yellow}\\
\hline
\end{tabular}\renewcommand{\arraystretch}{1.5}
&\renewcommand{\arraystretch}{1.5}
\begin{tabular}{|M|M|M|M|M|}
\hline
$\rightarrow$ &$0$&$+$&$-$&$\pm$\\
\hline
$0$&\cellcolor[gray]{0}&\cellcolor[gray]{0}&\cellcolor[gray]{0}&\cellcolor[gray]{0}\\
\hline
$+$&5&6&7&\cellcolor{Yellow}\\
\hline
$-$&\cellcolor[gray]{0}&\cellcolor[gray]{0}&\cellcolor[gray]{0}&\cellcolor[gray]{0}\\
\hline
$\pm$&\cellcolor{Orange}&\cellcolor{Yellow}&\cellcolor{Yellow}&\cellcolor{Red}\\
\hline
\end{tabular}\renewcommand{\arraystretch}{1.5}
&\renewcommand{\arraystretch}{1.5}
\begin{tabular}{|M|M|M|M|M|}
\hline
$\rightarrow$ &$0$&$+$&$-$&$\pm$\\
\hline
$0$&\cellcolor[gray]{0}&\cellcolor[gray]{0}&\cellcolor[gray]{0}&\cellcolor[gray]{0}\\
\hline
$+$&7&5&7&\cellcolor{Orange}\\
\hline
$-$&\cellcolor[gray]{0}&\cellcolor[gray]{0}&\cellcolor[gray]{0}&\cellcolor[gray]{0}\\
\hline
$\pm$&\cellcolor{Yellow}&\cellcolor{Yellow}&\cellcolor{Yellow}&\cellcolor[gray]{0}\\
\hline
\end{tabular}\renewcommand{\arraystretch}{1.5}
&\renewcommand{\arraystretch}{1.5}
\begin{tabular}{|M|M|M|M|M|}
\hline
$\rightarrow$ &$0$&$+$&$-$&$\pm$\\
\hline
$0$&\cellcolor[gray]{0}&\cellcolor[gray]{0}&\cellcolor[gray]{0}&\cellcolor[gray]{0}\\
\hline
$+$&7&6&7&\cellcolor{Red}\\
\hline
$-$&\cellcolor[gray]{0}&\cellcolor[gray]{0}&\cellcolor[gray]{0}&\cellcolor[gray]{0}\\
\hline
$\pm$&\cellcolor{Yellow}&\cellcolor{Orange}&\cellcolor{Yellow}&\cellcolor{Red}\\
\hline
\end{tabular}\renewcommand{\arraystretch}{1.5}
&\renewcommand{\arraystretch}{1.5}
\begin{tabular}{|M|M|M|M|M|}
\hline
$\rightarrow$ &$0$&$+$&$-$&$\pm$\\
\hline
$0$&\cellcolor[gray]{0}&\cellcolor[gray]{0}&\cellcolor[gray]{0}&\cellcolor[gray]{0}\\
\hline
$+$&7&\cellcolor{Yellow}&7&4\\
\hline
$-$&\cellcolor[gray]{0}&\cellcolor[gray]{0}&\cellcolor[gray]{0}&\cellcolor[gray]{0}\\
\hline
$\pm$&\cellcolor{Yellow}&\cellcolor{Red}&\cellcolor{Yellow}&\cellcolor{Orange}\\
\hline
\end{tabular}\renewcommand{\arraystretch}{1.5}
&\renewcommand{\arraystretch}{1.5}
\begin{tabular}{|M|M|M|M|M|}
\hline
$\rightarrow$ &$0$&$+$&$-$&$\pm$\\
\hline
$0$&\cellcolor[gray]{0}&\cellcolor[gray]{0}&\cellcolor[gray]{0}&\cellcolor[gray]{0}\\
\hline
$+$&\cellcolor[gray]{0}&7&7&7\\
\hline
$-$&\cellcolor[gray]{0}&\cellcolor[gray]{0}&\cellcolor[gray]{0}&\cellcolor[gray]{0}\\
\hline
$\pm$&\cellcolor{Yellow}&\cellcolor[gray]{0}&\cellcolor{Yellow}&\cellcolor{Yellow}\\
\hline
\end{tabular}\renewcommand{\arraystretch}{1.5}
&\renewcommand{\arraystretch}{1.5}
\begin{tabular}{|M|M|M|M|M|}
\hline
$\rightarrow$ &$0$&$+$&$-$&$\pm$\\
\hline
$0$&\cellcolor[gray]{0}&\cellcolor[gray]{0}&\cellcolor[gray]{0}&\cellcolor[gray]{0}\\
\hline
$+$&7&\cellcolor[gray]{0}&7&7\\
\hline
$-$&\cellcolor[gray]{0}&\cellcolor[gray]{0}&\cellcolor[gray]{0}&\cellcolor[gray]{0}\\
\hline
$\pm$&\cellcolor{Yellow}&\cellcolor{Yellow}&\cellcolor{Yellow}&\cellcolor[gray]{0}\\
\hline
\end{tabular}\renewcommand{\arraystretch}{1.5}

\end{tabular}
	}}
	\label{table:implicationalVGPSzondi}
\end{table}
We continue to define what we mean by 
	our simple implicational invariants announced in the introduction.
As announced there, 
	these invariants are ground instances of intuitionistic implication, by which 
		we mean that they are of the visually tractable, diagrammatic form $A\limp A'$ rather than 
			being of the more general, not generally visually tractable form $\phi\limp\phi'$.
As an example of what we mean by visually tractable, diagrammatic form,
	consider Table~\ref{table:implicationalVGPSzondi}.
For a given SPP-sequence $P$, 
	we postulate the algorithmically extracted set $\invs(P)$ of these invariants that 
		hold throughout $P$ (see Definition~\ref{definition:Invariants}) as the axioms of 
			the personality theory $\Clo{}{}(\invs(P))$ that we associate with $P$.
These axioms thus capture 
	those \emph{logical dependencies} between 
		signed factors that are invariant in $P$ in the sense of holding throughout $P$.
The algorithm for this axiom extraction and visualisation is displayed in Listing~\ref{Listing:Algorithm} and
	will be explained shortly.
Note that 
	given that these invariants hold throughout a sequence that
		has been generated by an iterated procedure, that is, an iterated execution of the Szondi-test, 
			they can also be understood as \emph{loop invariants,} which is 
				a core concept in the science of computer programming \cite{LoopInvariants}.
So our algorithm for finding psychological invariants can actually also be understood and even be used as 
	a method for inferring loop invariants from 
		program execution traces in computer science.
\begin{definition}[Simple implicational invariants]\label{definition:Invariants}
Let the mapping $\leftI:\SPP\to\statements(\atoms)$ be such that 
					$$\begin{array}{@{}r@{}}
						 \leftI(((\F{h}{,s_{1}}), (\F{s}{,s_{2}}), (\F{e}{,s_{3}}), (\F{hy}{,s_{4}}), 
			(\F{k}{,s_{5}}), (\F{p}{,s_{6}}), (\F{d}{,s_{7}}), (\F{m}{,s_{8}})))=\\
						\F{h}{s_{1}}\land\F{s}{s_{2}}\land\F{e}{s_{3}}\land\F{hy}{s_{4}}\land\F{k}{s_{5}}\land\F{p}{s_{6}}\land\F{d}{s_{7}}\land\F{m}{s_{8}}\,.
					\end{array}$$
	
	Then, 
		define the mapping $\invs:\SPPs\to2^{\statements(\atoms)}$ of \emph{simple implicational invariants} such that 
			for every $ P\in\SPPs$, 
				$$\invs( P):=
					\{\,A\limp A'\mid
						\text{\begin{tabular}{@{\ }c@{\ }}
								for every $ P'\in\SPPs$, if $P\sqsubseteq P'$\\ 
								then $\leftI(\pi_{1}( P'))\vdash_{\emptyset} A\limp A'$
							\end{tabular}}\,\}\,,$$
		where $\pi_{1}:\SPPs\to\SPP$ is projection onto the first SPP component.
\end{definition}
\noindent
Notice the three implications ``if---then,'' $\vdash$, and $\limp$ of different logical level, and 
note that we use ``if---then'' and ``implies'' synonymously.
This definition can be cast into an algorithm of linear complexity in the length of $P$, 
	for example as described by 
		the Java-program displayed in Listing~\ref{Listing:Algorithm}, and
			the result $\invs(P)$ of its computation diagrammatically displayed as in Table~\ref{table:implicationalVGPSzondi}.
\begin{lstlisting}[float,caption={Update algorithm},label=Listing:Algorithm]
public void update (Vector<Signature[]> profiles) { 
	// 1. CALCULATION OF MATERIAL IMPLICATIONS
	// in the first profile
	Signature[] fstp = profiles.firstElement();
	// consequent-oriented processing (consequent loop),  
	// round-robin treatment of each factor as consequent
	for (int c=0; c<8; c++) { 
		Signature cmodq = moduloQuanta(fstp[c]);
		// 1.1 EVERYTHING IMPLIES TRUTH (antecedent loop), 
		// round-robin treatment of each factor as antecedent
		for (int a=0; a<8; a++) {  
			// signature-value loop	
			for (int v=0; v<4; v++) { 
				// discount corresponding table-cell value 
				(factors[a][c]).signatures[v][code(cmodq)]--; 
			}
		}
		// 1.2 FALSEHOOD IMPLIES EVERYTHING---ALSO FALSEHOOD; 
		// everything is: all other consequent signatures
		for (Signature cc : coSet(cmodq)) { 
			// round-robin treatment of each factor as antecedent 
			for (int a=0; a<8; a++) { 
				// false is: all other antecedent signatures
				for (Signature ca : coSet(moduloQuanta(fstp[a]))) { 
					// discount corresponding table-cell value 
					(factors[a][c]).signatures[code(ca)][code(cc)]--; 
				}
			}
		}
	}
	// 2. CALCULATION OF INTUITIONISTIC IMPLICATIONS:
	// forward invariance of material implications
	if (profiles.size()>1) {
		// garbage-collect the processed profile
		profiles.remove(0); 
		// recursively descend on the remaining profiles
		update(profiles);
	}
}
\end{lstlisting}
The on-line Szondi-test \cite{SzondiTestWebApp} also uses this program as a subroutine.
Lines starting with ``\verb|//|\,'' are comments.
Notice that every loop in the program has fixed complexity, and 
	that we simply process the head of $P$---the first profile in $P$---in Line~2--30 and then 
		recur on the tail of the remaining profiles in $P$ in Line~31--38.
The loop-nesting depth is four.
The program updates 
	a table---called \emph{factors} in Listing~\ref{Listing:Algorithm}---of eight times eight subtables---called \emph{signatures} in Listing~\ref{Listing:Algorithm}---of four times four content cells 
		as displayed in Table~\ref{table:implicationalVGPSzondi}, 
			each of whose cells is initialised with a value equal to the length of $P$ (e.g., 10).
To update this table means to
	discount the initial value of its cells according to the following strategy  
		inspired by Kripke's model-theoretic interpretation of intuitionistic implication as 
			forward invariance of material implication \cite{KripkeSemanticsIL,sep-logic-intuitionistic} and  
				adapted to our setting in Definition~\ref{definition:Invariants}:
	\begin{enumerate}
		\item Calculate all material implications in the first profile in $P$, 
				called \emph{profiles} in Listing~\ref{Listing:Algorithm}, according to
					the definition of material implication recalled in Table~\ref{table:MaterialImplication}.
					\begin{table}[t]
						\centering
						\caption{Material implication $\supset$}
						\medskip
						\begin{tabular}{r|c|c|c|}
							\cline{2-4}
							\multicolumn{1}{c|}{} & $A$ & $A'$ & $A\supset A'$\\
							\cline{2-4}
							1. & false & false & true\\
							2. & false & true & true\\
							3. & true & false & false\\
							4. & true & true & true\\
							\cline{2-4}
						\end{tabular}
						\label{table:MaterialImplication}
					\end{table}
					There, Line~1, 2, and 4 can be summarised by
						the slogan ``Everything implies truth'' and 
						Line~1 and 2 by the slogan ``Falsehood implies everything.''
					In Listing~\ref{Listing:Algorithm},
						these well-known slogans correspond to the meaning of our code in 
						Line~9--17 and Line~18--29, respectively.
					There, 
						the function \emph{moduloQuanta} simply returns 
							its argument signature without quanta for graphical tractability,  
						the function \emph{code}
							the subtable line number of its argument, and
						the function \emph{coSet} 
							the set of all plain signatures (those signatures without quanta) minus
								the argument signature. 
					For example, applying 
						\begin{itemize}
							\item \emph{moduloQuanta} to the signature $\pbbb$ 
									returns the signature $\p$,  
							\item \emph{code} to the signature $\p$ 
									returns the line number 1, and
							\item \emph{coSet} to the signature $\p$ 
									returns the set of signatures $\{\m,\n,\pm\}$.
						\end{itemize}
		\item Then calculate those material implications that are actually even intuitionistic implications by 
				recurring on the tail of $P$.
	\end{enumerate}
On termination of program execution,
	each table cell that corresponds to an intuitionistic implication will
		contain the number 0 and be painted black.
Cells containing the number 1 
	will be painted red (and correspond to intuitionistic implications of the tail of $P$),
		those containing the number 2 will be painted orange, and 
		those containing the number 3 yellow.
Table cells containing other numbers 
	will not be painted for lack of relevance and thus 
	will just display the number of missing discounts as 
		distance to count as intuitionistic implications.\label{page:DistanceToInvariance}

Observe in Table~\ref{table:implicationalVGPSzondi} that
	the whole diagonal from the top left corner down to the bottom right corner is painted black.
This state of affairs reflects   
	the reflexivity property $\vdash_{\Gamma}\phi\limp\phi$ of intuitionistic implication.
Similarly,
	if both 
		a cell representing some formula $A\limp A'$ as well as  
		another cell representing some formula $A'\limp A''$
	are painted black 
	then the cell representing the formula $A\limp A''$ will also be painted black.
Consider the following four example triples:
		\begin{itemize}
			\item $\F{e}{\p}\limp\F{s}{\n}$, $\F{s}{\n}\limp\F{k}{\pm}$, and $\F{e}{\p}\limp\F{k}{\pm}$\,;
			\item $\F{k}{\p}\limp\F{s}{\m}$, $\F{s}{\m}\limp\F{p}{\p}$, and $\F{k}{\p}\limp\F{p}{\p}$\,;
			\item $\F{p}{\pm}\limp\F{s}{\n}$, $\F{s}{\n}\limp\F{k}{\pm}$, and $\F{p}{\pm}\limp\F{k}{\pm}$\,;
			\item $\F{m}{\pm}\limp\F{d}{\p}$, $\F{d}{\p}\limp\F{hy}{\pm}$, and $\F{m}{\pm}\limp\F{hy}{\pm}$\,.
		\end{itemize}
This state of affairs reflects 
	the transitivity property of intuitionistic implication, which for general formulas $\phi$ and $\phi'$ is:
		$$\text{if $\vdash_{\Gamma}\phi\limp\phi'$ and $\vdash_{\Gamma}\phi'\limp\phi''$ 
				then $\vdash_{\Gamma}\phi\limp\phi''$.}$$
Of course, 
	reflexivity and transitivity are two logical properties, which
		will show up in the diagram of any $\invs(P)$.
(In so far,
	these properties also reflect an axiomatic redundancy of $\invs(P)$, which
		however is not our concern here.)

In contrast,
	the following properties displayed in 
		Table~\ref{table:implicationalVGPSzondi} are \emph{psychological} in that 
			they are proper to the personality profile displayed in Table~\ref{table:VGPSzondi}, 
				from which they have been extracted, namely: 
\begin{description}
	\item[Vacuous implications] This class of intuitionistic implications is 
			visually characterised by a horizontal or vertical line of black cells 
			throughout the whole diagram width and diagram height, respectively.
			In the diagram displayed in Table~\ref{table:implicationalVGPSzondi},
				there is a single vertical line of such implications, which
					says:
						$$\text{Whenever something is true then $\F{h}{\m}$ is true.}$$
						This simply holds because $\F{h}{\m}$ is true throughout the whole 
						test result in Table~\ref{table:VGPSzondi}, and 
						thus the logical slogan ``Everything implies truth'' applies.
						
			The other, that is, the horizontal lines of vacuous intuitionistic implications hold due to 
				the other logical slogan ``Falsehood implies everything.''
	\item[Non-vacuous implications] This class of intuitionistic implications is 
			visually characterised by isolated black cells. They are 
			the psychologically truly interesting implications.
			They are, from the top left to the bottom right of 
			the diagram in Table~\ref{table:implicationalVGPSzondi}, and
			together with their rough psychological meaning in Szondi's system 
			(recall Table~\ref{table:SzondiFactors} and, if need be,  
				consult \cite{Szondi:ETD:Band1}):
			\begin{enumerate}
				\item $\F{s}{\n}\limp\F{k}{\pm}$\,. 
						Whenever the testee is inactive (externally)   
						then he has internal compulsive behaviour 
						(e.g., is experiencing a dilemma).
						See for example the below Item~3 and 6, where
							these two implicationally related reactions also appear conjunctively.
				\item $\F{s}{\m}\limp\F{p}{\p}$\,.
						Whenever the testee is receptively passive (e.g., masochism) 
						then he wants to be more than he actually is (e.g., megalomania).
						See for example the below Item~5, where
							these two implicationally related reactions also appear conjunctively.
				\item $\F{e}{\p}\limp(\F{s}{\n}\land\F{hy}{\pm}\land\F{k}{\pm}\land\F{p}{\p}\land\F{d}{\n}\land\F{m}{\p}):$ Whenever the testee has ethical behaviour then he 
					\begin{enumerate}
						\item is inactive ($\F{s}{\n}$). 
							That is, inactivity is a necessary condition for the testee's ethical behaviour, and 
								thus his ethical behaviour is in a behavioural (not logical) sense vacuous.
						\item is morally ambivalent ($\F{hy}{\pm}$). Thus the testee's ethical behaviour need not be moral. Indeed, inactivity need not be moral.
						\item has internal compulsive behaviour ($\F{k}{\pm}$). 
								Maybe the testee's inactivity is due to an internally experienced dilemma?
						\item wants to be more than he is ($\F{p}{\p}$).
								The testee's inactivity may not be conducive to the fulfilment of his desire, but
									his desire may well be co-determined by his inactivity.
						\item is faithfully indifferent ($\F{d}{\n}$). Indeed, faithfulness (in a general sense) and ethics may be experienced as orthogonal issues.
						\item approves of bindings in his relationships ($\F{m}{\p}$). Thus for the testee,
								bindings but not necessarily their faithfulness are ethical.
					\end{enumerate}
				\item $\F{hy}{\p}\limp(\F{d}{\n}\land\F{m}{\p}):$ Whenever the testee has immoral behaviour then he 
					\begin{enumerate}
						\item is faithfully indifferent ($\F{d}{\n}$). 
						\item approves of bindings in his relationships ($\F{m}{\p}$). 
					\end{enumerate}
					Thus the 
						testee's faithfulness indifference as well as 
						his binding attitude is stable with respect to ethics and immorality.
				\item $\F{k}{\p}\limp(\F{s}{\m}\land\F{p}{\p}):$ Whenever the testee wants to have more than he has then he 
					\begin{enumerate}
						\item is receptively passive ($\F{s}{\m}$).
						\item wants to be more than he is ($\F{p}{\p}$).
					\end{enumerate}
					Again, the testee's receptive passivity may not be conducive to the fulfilment of his desires, but his desires may well be co-determined by his receptive passivity.
				\item $\F{p}{\pm}\limp(\F{s}{\n}\land\F{k}{\pm}):$ Whenever the testee is ambivalent with respect to being more or less than he is then he 
					\begin{enumerate}
						\item is inactive ($\F{s}{\n}$). 
								The testee's ambivalence may well be a deeper dilemma that 
									is the cause of his activity blockage.
						\item has internal compulsive behaviour ($\F{k}{\pm}$).
								This could be a confirmation of the testee's suspected dilemma.
					\end{enumerate}
				\item $\F{d}{\n}\limp\F{m}{\p}$\,. Whenever the testee is faithfully indifferent 
						then he approves of bindings in his relationships.
							See for example the above Item~3 and 4, where 
								these two implicationally related reactions also appear conjunctively.
				\item $\F{d}{\p}\limp\F{hy}{\pm}$\,. Whenever the testee is unfaithful then he is morally ambivalent. See for example the below Item~10, where 
								these two implicationally related reactions also appear conjunctively.
				\item $\F{m}{\p}\limp\F{d}{\n}$\,. Whenever the testee approves of bindings in his relationships then he is faithfully indifferent. For example see Item~3 and 4, where 
								these two implicationally related reactions also appear conjunctively.
				\item $\F{m}{\pm}\limp(\F{hy}{\pm}\land\F{d}{\p}):$ Whenever the testee is ambivalent in his  attitude towards bindings in his relationships then he 
					\begin{enumerate}
						\item is morally ambivalent ($\F{hy}{\pm}$).
						\item is unfaithful ($\F{d}{\p}$).
					\end{enumerate}
			\end{enumerate}
\end{description}
Observe that from the above invariants in the given $P$ we can deduce that: 
	\begin{align*}
		&\vdash_{\invs(P)}(\F{e}{\p}\lor\F{s}{\n}\lor\F{p}{\pm})\limp\F{k}{\pm}\\
		&\vdash_{\invs(P)}(\F{e}{\p}\lor\F{s}{\m}\lor\F{k}{\p})\limp\F{p}{\p}\\
		&\vdash_{\invs(P)}(\F{e}{\p}\lor\F{hy}{\p})\limp(\F{d}{\n}\land\F{m}{\p})\\
		&\vdash_{\invs(P)}(\F{e}{\p}\lor\F{hy}{\p}\lor\F{m}{\p})\limp\F{d}{\n}\\
		&\vdash_{\invs(P)}(\F{e}{\p}\lor\F{hy}{\p}\lor\F{d}{\n})\limp\F{m}{\p}\\
		&\vdash_{\invs(P)}(\F{e}{\p}\lor\F{p}{\pm})\limp\F{s}{\n}\\
		&\vdash_{\invs(P)}(\F{e}{\p}\lor\F{d}{\p}\lor\F{m}{\pm})\limp\F{hy}{\pm}
	\end{align*}
From our diagrammatic reasonings, 
	it becomes clear that 
		the signed factor $\F{e}{\p}$ and to a lesser extent the signed factor $\F{hy}{\p}$ are 
			the two most important \emph{causal factors} in $P$---and 
				thus for the testee represented by $P$---in the following sense: 
			\begin{enumerate}
				\item these factors are non-vacuously implied by no other signed factor, but
				\item they individually and non-vacuously imply most other signed factors.
			\end{enumerate}
Thus the testee's personality is determined to a large extent by these two signed factors,
	in spite of the fact that they only occur in $P$ once and twice, respectively!

\paragraph{Diagrammatic reasoning in a couple}
Given two (or more) SPP-sequences $P$ and $P'$,
	we can compute their axiom bases $\invs(P)$ and $\invs(P')$, 
		and visualise them as diagrams.
We can then graphically compute the join and the meet of $P$ and $P'$, that is, 
	the union and the intersection of $\invs(P)$ and $\invs(P')$, respectively, by simply
		superposing the two diagrams as printed on overhead-projector foils, and then  
		adding their cells and 
		pinpointing their common cells on a third and fourth superimposed foil, respectively.
Of course, this graphical computation can instead also be programmed on a computer 
	(e.g., for studying groups).

\paragraph{Conjunctive implicational invariants}
As indicated,
	our algorithmically extracted implicational invariants $A\limp A'$ are simple in that they have 
		a single atomic antecedent $A$ and
		a single atomic consequent $A'$.
An interesting generalisation of these simple implicational invariants is 
	to allow finite conjunctions $A_{1}\land\ldots\land A_{n}$ of 
		atomic formulas $A_{1},\ldots,A_{n}\in\atoms$ as antecedents.
This generalisation, though graphically not generally tractable in two dimensions, is interesting because with it,  
	signed personality factors, represented by atomic formulas, 
		can be analysed in terms of their 
			\emph{individually necessary and jointly sufficient conditions,} and 
				thus \emph{logically characterised in terms of each other.}
More precisely, 
	we mean by this characterisation that from 
		\begin{enumerate}
			\item $\vdash_{\invs(P)}(A_{1}\land\ldots\land A_{n})\limp A'$, that is,
				the atomic formulas $A_{1},\ldots,A_{n}$ are \emph{jointly sufficient conditions} for the atomic formula $A'$, and 
			\item $\vdash_{\invs(P)}(A'\limp A_{1})\land\ldots\land(A'\limp A_{n})$, that is,
				the atomic formulas $A_{1},\ldots,A_{n}$ are \emph{individually necessary conditions} for the atomic formula $A'$,
		\end{enumerate}
	we can deduce the following \emph{equivalence characterisation} of $A':$ 
		$$\vdash_{\invs(P)}(A_{1}\land\ldots\land A_{n})\leftrightarrow A'.$$ 
Obviously, the truth of Item~2 can be ascertained graphically with our automatic procedure, and 
Item~1 can be ascertained interactively with the following semi-automatic procedure, involving
	a standard, efficient database query language:
\begin{enumerate}
	\item Transcribe the given $P$ (e.g., Table~\ref{table:VGPSzondi}) into \emph{non-recursive Datalog} \cite{FoundationsOfDatabases};
	\item Formulate and then query the resulting database with the jointly sufficient conditions that
			you suspect to be true.
\end{enumerate}

\begin{proposition}[Suffix closure of $\invs$]\label{proposition:invs} 
	For every $P,P'\in\SPPs$,
	\begin{enumerate} 
		\item $\invs(P\conc P')\subseteq\invs(P')$
		\item $P\sqsubseteq P'$ implies $\invs(P)\subseteq\invs(P')$
	\end{enumerate}
\end{proposition}
\begin{proof}
	For (1),
		consider:
	\begin{enumerate}
		\item\quad $P,P'\in\SPPs$\hfill hypothesis\\[-1.25\baselineskip]
		\item\qquad $\phi\in\invs(P\conc P')$\hfill hypothesis\\[-1.25\baselineskip]
		\item\qquad \begin{tabular}{@{}l@{}}
					there are $A,A'\in\atoms$ such that $\phi=A\limp A'$ and\\ 
					for every $ P''\in\SPPs$, 
								$P\conc P'\sqsubseteq P''$ implies $\leftI(\pi_{1}( P''))\vdash \phi$
					\end{tabular}\hfill 2\\[-0.75\baselineskip]
		\item\qquad\quad \begin{tabular}{@{}l@{}} 
					$\phi=A\limp A'$ and for every $ P''\in\SPPs$,\\  
						$P\conc P'\sqsubseteq P''$ implies $\leftI(\pi_{1}( P''))\vdash \phi$
					\end{tabular}\hfill hypothesis\\[-0.75\baselineskip]
		\item\qquad\qquad $P''\in\SPPs$\hfill hypothesis\\[-1.25\baselineskip]
		\item\qquad\qquad\quad $P'\sqsubseteq P''$\hfill hypothesis\\[-1.25\baselineskip]
		\item\qquad\qquad\quad $P\conc P'\sqsubseteq P''$\hfill 6, Fact~\ref{fact:conc}\\[-1.25\baselineskip]
		\item\qquad\qquad\quad $\leftI(\pi_{1}( P''))\vdash \phi$\hfill 4, 5, 7\\[-1.25\baselineskip]
		\item\qquad\qquad $P'\sqsubseteq P''$ implies $\leftI(\pi_{1}( P''))\vdash \phi$\hfill 6--8\\[-1.25\baselineskip]
		\item\qquad\quad for every $ P''\in\SPPs$,  
						$P'\sqsubseteq P''$ implies $\leftI(\pi_{1}( P''))\vdash \phi$\hfill 5--9\\[-1.25\baselineskip]
		\item\qquad\quad \begin{tabular}{@{}l@{}} 
					$\phi=A\limp A'$ and for every $ P''\in\SPPs$,\\  
						$P'\sqsubseteq P''$ implies $\leftI(\pi_{1}( P''))\vdash \phi$
					\end{tabular}\hfill 4, 10\\[-0.75\baselineskip]
		\item\qquad\quad \begin{tabular}{@{}l@{}}
					there are $A,A'\in\atoms$ such that $\phi=A\limp A'$ and\\ 
					for every $ P''\in\SPPs$, 
								$P'\sqsubseteq P''$ implies $\leftI(\pi_{1}( P''))\vdash \phi$
					\end{tabular}\hfill 11\\[-0.75\baselineskip]
		\item\qquad\quad $\phi\in\invs(P')$\hfill 12\\[-1.25\baselineskip]
		\item\qquad $\phi\in\invs(P')$\hfill 3, 4--13\\[-1.25\baselineskip]
		\item\quad $\invs(P\conc P')\subseteq\invs(P')$\hfill 2--14\\[-1.25\baselineskip]
		\item for every $P,P'\in\SPPs$, $\invs(P\conc P')\subseteq\invs(P')$\hfill 1--15.
	\end{enumerate}
For (2),
	consider:
	\begin{enumerate}
		\item\quad $P,P'\in\SPPs$\hfill hypothesis\\[-1.25\baselineskip]
		\item\qquad $P\sqsubseteq P'$\hfill hypothesis\\[-1.25\baselineskip]
		\item\qquad $P=P'$ or there is $P''\in\SPPs$ such that $P=P''\conc P'$\hfill 2\\[-1.25\baselineskip]
		\item\qquad $P=P'$ implies $\leftG{\{P\}}\subseteq\leftG{\{P'\}}$\hfill equality law\\[-1.25\baselineskip]		
		\item\qquad\quad there is $P''\in\SPPs$ such that $P=P''\conc P'$\hfill hypothesis\\[-1.25\baselineskip]		
		\item\qquad\qquad $P''\in\SPPs$ and $P=P''\conc P'$\hfill hypothesis\\[-1.25\baselineskip]
		\item\qquad\qquad\quad $\phi\in\invs(P)$\hfill hypothesis\\[-1.25\baselineskip]
		\item\qquad\qquad\quad $\phi\in\invs(P''\conc P')$\hfill 6, 7\\[-1.25\baselineskip]
		\item\qquad\qquad\quad $\invs(P''\conc P')\subseteq\invs(P')$\hfill 1, Propostion~\ref{proposition:invs}.1\\[-1.25\baselineskip]
		\item\qquad\qquad\quad $\phi\in\invs(P')$\hfill 8, 9\\[-1.25\baselineskip]
		\item\qquad\qquad $\invs(P)\subseteq\invs(P')$\hfill 7--10\\[-1.25\baselineskip]
		\item\qquad\quad $\invs(P)\subseteq\invs(P')$\hfill 5, 6--11\\[-1.25\baselineskip]
		\item\qquad \begin{tabular}{@{}l@{}}
					there is $P''\in\SPPs$ such that $P=P''\conc P'$\\
					implies $\invs(P)\subseteq\invs(P')$
					\end{tabular}\hfill 5--12\\[-0.75\baselineskip]
		\item\qquad $\invs(P)\subseteq\invs(P')$\hfill 3, 4, 13\\[-1.25\baselineskip]
		\item\quad $P\sqsubseteq P'$ implies $\invs(P)\subseteq\invs(P')$\hfill 2--14\\[-1.25\baselineskip]
		\item for every $P,P'\in\SPPs$, $P\sqsubseteq P'$ implies $\invs(P)\subseteq\invs(P')$\hfill 1--15.\\[-1.25\baselineskip]
	\end{enumerate}
\end{proof}

\begin{proposition}\ 
	\begin{enumerate}
		\item $\vdash_{\invs(P\conc P')}\phi$ implies $\vdash_{\invs(P')}\phi$
		\item If $P\sqsubseteq P'$ and $\vdash_{\invs(P)}\phi$ then $\vdash_{\invs(P')}\phi$\,.
	\end{enumerate}
\end{proposition}
\begin{proof}
	Combine 
		Fact~\ref{fact:ClosureOperator}.2 with 
			Proposition~\ref{proposition:invs}.1 and
			Proposition~\ref{proposition:invs}.2, respectively.
\end{proof}

The following property means that 
	our personality theories have  
		the desired prime-filter property (see Proposition~\ref{proposition:SpecialCase}), 
			as announced in the introduction.
\begin{theorem}[Disjunction Property]\label{theorem:DisjunctionProperty}
	$$\text{If $\vdash_{\invs(P)}\phi\lor\phi'$ then $\vdash_{\invs(P)}\phi$ or $\vdash_{\invs(P)}\phi'$.}$$
\end{theorem}
\begin{proof}
	Our proof strategy is 
		to adapt de Jongh's strategy in \cite{DeJonghOnDP} to our simpler setting, thanks to which  
			our proof reduces to G\"odel's proof of the disjunction property of 
				a basic intuitionistic theory \cite{Goedel:OnIPC} such as our $\Clo{}{}(\emptyset)$:
	So suppose that $\vdash_{\invs(P)}\phi\lor\phi'$.
	Adapting an observation from \cite{DeJonghOnDP},
		we can assert that 
			$\vdash_{\invs(P)}\phi\lor\phi'$ if and only if 
			$\vdash_{\emptyset}\bigwedge\invs(P)\limp(\phi\lor\phi')$.
	Thus $\vdash_{\emptyset}\bigwedge\invs(P)\limp(\phi\lor\phi')$.
	Hence $\vdash_{\emptyset}(\bigwedge\invs(P)\limp\phi)\lor(\bigwedge\invs(P)\limp\phi')$.
	Hence 
		$\vdash_{\emptyset}(\bigwedge\invs(P)\limp\phi)$ or 
		$\vdash_{\emptyset}(\bigwedge\invs(P)\limp\phi')$ by G\"odel's proof.
	Hence 
		$\vdash_{\invs(P)}\phi$ or 
		$\vdash_{\invs(P)}\phi'$ again by de Jongh's observation.
\end{proof}

\section{Personality categorisation}\label{section:Categorisation}
In this section,
	we present the part of our framework for 
		the mathematical categorisation of 
			personality axioms into formal personality theories, as  
				these axioms might have been discovered with 
					the methodology presented in the previous section.
As announced in the introduction,
	personality theories and personality-test data are related by a Galois-connection 
		\cite[Chapter~7]{DaveyPriestley}.
We start with defining this connection and the two personality (powerset) spaces that it connects.

\begin{definition}[Personality algebras]\label{definition:PersonalityAlgebras}
	Let the mappings 
		$\rightG{}:2^{\statements(\atoms)}\to2^{\SPPs}$, called \emph{right polarity,} and 
		$\leftG{}:2^{\SPPs}\to2^{\statements(\atoms)}$, called \emph{left polarity,} be such that 
	\begin{itemize}
		\item $\rightG{\Phi}:=\{\,P\in\SPPs\mid\text{for every $\phi\in\Phi$, $\vdash_{\invs(P)}\phi$}\,\}$ and 
		\item $\leftG{\mathcal{P}}:=\{\,\phi\in\statements(\atoms)\mid\text{for every $P\in\mathcal{P}$, $\vdash_{\invs(P)}\phi$}\,\}\,.$
	\end{itemize}
	Further let 
		${\equiv}\subseteq2^{\statements(\atoms)}\times2^{\statements(\atoms)}$ and 
		${\equiv}\subseteq2^{\SPPs}\times2^{\SPPs}$ be their \emph{kernels,} that is, 
		for every $\Phi,\Phi'\in2^{\statements(\atoms)}$, 
			$\Phi\equiv\Phi'$ by definition if and only if $\rightG{\Phi}=\rightG{\Phi'}$ and 
		for every $\mathcal{P},\mathcal{P}'\in2^{\SPPs}$, 
			$\mathcal{P}\equiv\mathcal{P}'$ by definition if and only if $\leftG{\mathcal{P}}=\leftG{\mathcal{P}'}$,
		respectively.
	
Then, 
	for each one of the two (inclusion-ordered, Boolean) powerset algebras 
	$$\text{
		$\langle\,2^{\statements(\atoms)},\emptyset,\cap,\cup,\statements(\atoms),\overline{\,\cdot\,},\subseteq\,\rangle\, 
		\autorightleftharpoons{$_\triangleright$}{$^\triangleleft$}\ 
		\langle\,2^{\SPPs},\emptyset,\cap,\cup,\SPPs,\overline{\,\cdot\,},\subseteq\,\rangle$}\,,$$
define its \emph{(ordered) quotient join semi-lattice with bottom} 
	(and thus idempotent commutative monoid) modulo its kernel as in Table~\ref{table:LTAlgebras}.
			\begin{table}[t]
			\centering
			\caption{Quotient algebras}
			\smallskip
			\resizebox{\textwidth}{!}{$\begin{array}{|r@{\ \ }r@{\ \ }l|r@{\ \ }r@{\ \ }l|}
				\hline
				\multicolumn{3}{|c|}{\text{Statements}} & 
				\multicolumn{3}{c|}{\text{Test results}}\\
				\hline
				&&&&&\\[-3\jot]
				\top &:=& [\statements(\atoms)]_{\equiv} & 
					\top &:=& [\SPPs]_{\equiv}\\[\jot]
				[\Phi]_{\equiv}\sqcup[\Phi']_{\equiv} &:=& [\Phi\cup\Phi']_{\equiv} & 
					[\mathcal{P}]_{\equiv}\sqcup[\mathcal{P}']_{\equiv} &:=& 
						[\mathcal{P}\cup\mathcal{P}']_{\equiv} \\[\jot]
				\bot &:=& [\emptyset]_{\equiv} & \bot &:=& [\emptyset]_{\equiv}\\[\jot]
				[\Phi]_{\equiv}\sqsubseteq[\Phi']_{\equiv} &\text{:iff}& 
					[\Phi]_{\equiv}\sqcup[\Phi']_{\equiv}=[\Phi']_{\equiv} &
				[\mathcal{P}]_{\equiv}\sqsubseteq[\mathcal{P}']_{\equiv} &\text{:iff}& 
					[\mathcal{P}]_{\equiv}\sqcup[\mathcal{P}']_{\equiv}=[\mathcal{P}']_{\equiv}\\[\jot]
				\hline
		\end{array}$}
		\label{table:LTAlgebras}
		\end{table}
\end{definition}
Note that our focus is on the powerset and not on the quotient algebras.
The purpose of the quotient algebras is simply to indicate the maximally 
	definable algebraic structure in our context.
As a matter of fact,
	only the join- but not the meet-operation is well-defined in the quotient algebra  
		(see Corollary~\ref{corollary:WellDefinedness}).

\begin{proposition}[Basic properties of personality theories]\label{proposition:SpecialCase}\ 
	\begin{enumerate}
		\item $\leftG{\{P\}}=(\Clo{}{}\circ\invs)(P)$\quad(generalisation to sets)
		\item $P\sqsubseteq P'$ implies $\leftG{\{P\}}\subseteq\leftG{\{P'\}}$\quad(monotonicity)
		\item prime filter properties:
				\begin{enumerate}
					\item if 
							$\phi\in\leftG{\{P\}}$ and 
							$\phi'\in\leftG{\{P\}}$ then 
							$\phi\land\phi'\in\leftG{\{P\}}$ (and vice versa)
					\item if 
							$\phi\in\leftG{\{P\}}$ and 
							$\phi'\in\statements(\atoms)$ and 
							$\phi\vdash_{\invs(P)}\phi'$ then 
							$\phi'\in\leftG{\{P\}}$
					\item if $\phi\lor\phi'\in\leftG{\{P\}}$ then 
							$\phi\in\leftG{\{P\}}$ or $\phi'\in\leftG{\{P\}}$ (and vice versa)
					\end{enumerate}
					($\leftG{\{P\}}$ is an intuitionistic theory.)
		\item for every $\phi,\phi',\phi''\in\statements(\atoms)$, 
				$$\text{if $\phi\lor\phi',\phi\lor\phi''\in\bigcap_{P\in\mathcal{P}}\leftG{\{P\}}$ 
						then $\phi\lor(\phi'\land\phi'')\in\bigcap_{P\in\mathcal{P}}\leftG{\{P\}}$}$$\quad($\bigcap_{P\in\mathcal{P}}\leftG{\{P\}}$ is a distributive filter.)
	\end{enumerate}
\end{proposition}
\begin{proof}
For (1), consider that 
$$\begin{array}{r@{\ \ }c@{\ \ }l}
	\leftG{\{P\}} &=& 
		\{\,\phi\in\statements(\atoms)\,\mid
			\text{for every $P'\in\{P\}$, $\vdash_{\invs(P')}\phi$}\,\}\\[\jot]
			&=& \{\,\phi\in\statements(\atoms)\,\mid\ \vdash_{\invs(P)}\phi\,\}\\[\jot]
			&=& \{\,\phi\in\statements(\atoms)\,\mid\phi\in\Clo{}{}(\invs(P))\,\}\\[\jot]
			&=& \Clo{}{}(\invs(P))\\[\jot]
			&=& (\Clo{}{}\circ\invs)(P)
\end{array}$$
	For (2), suppose that $P\sqsubseteq P'$.
	Hence $\invs(P)\subseteq\invs(P')$ by Proposition~\ref{proposition:invs}.2.
	Hence $\Clo{}{}(\invs(P))\subseteq\Clo{}{}(\invs(P'))$ by Fact~\ref{fact:ClosureOperator}.2.
	Thus $\leftG{\{P\}}\subseteq\leftG{\{P'\}}$ by (1).
	(3.a) follows from the fact that
		$(\phi\limp(\phi'\limp(\phi\land\phi'))\in\Clo{}{}(\emptyset)$
			(and $((\phi\land\phi')\limp\phi),((\phi\land\phi')\limp\phi')\in\Clo{}{}(\emptyset)$), 
				Fact~\ref{fact:ClosureOperator}.2, and
				(1).
	For (3.b), inspect definitions, and
	for (3.c),
		Theorem~\ref{theorem:DisjunctionProperty}, definitions, and (1).
	For (4),
		consider (3) and 
		recall that 
			intersections of prime filters are distributive filters 
				\cite[Exercise~10.9]{DaveyPriestley}.
\end{proof}

Now note the two macro-definitions 
	$\rightleftG{}:=\rightG{}\circ\leftG{}$ and 
	$\leftrightG{}:=\leftG{}\circ\rightG{}$ with 
		$\circ$ being function composition, as usual (from right to left, as usual too).

\begin{lemma}[Some useful properties of $\rightG{}$ and $\leftG{}$]\label{lemma:Properties}\ 
	\begin{enumerate}
		\item if $\Phi\subseteq\Phi'$ then $\rightG{\Phi'}\subseteq\rightG{\Phi}$\quad(\;$\rightG{}$ is antitone)
		\item if $\mathcal{P}\subseteq\mathcal{P}'$ then $\leftG{\mathcal{P}'}\subseteq\leftG{\mathcal{P}}$\quad(\,$\leftG{}$ is antitone)
		\item $\mathcal{P}\subseteq\rightG{(\leftG{\mathcal{P}})}$\quad(\;$\rightleftG{}$ is extensive)
		\item $\Phi\subseteq\leftG{(\rightG{\Phi})}$\quad(\,$\leftrightG{}$ is extensive)
		\item $\leftG{(\rightG{(\leftG{\mathcal{P}})})}=\leftG{\mathcal{P}}$
		\item $\rightG{(\leftG{(\rightG{\Phi})})}=\rightG{\Phi}$
		\item $\rightG{(\leftG{(\rightG{(\leftG{\mathcal{P}})})})}=\rightG{(\leftG{\mathcal{P}})}$\quad
				(\,$\rightleftG{}$ is idempotent)
		\item $\leftG{(\rightG{(\leftG{(\rightG{\Phi})})})}=\leftG{(\rightG{\Phi})}$\quad
				(\,$\leftrightG{}$ is idempotent)
		\item if $\mathcal{P}\subseteq\mathcal{P}'$ 
				then $\rightG{(\leftG{\mathcal{P}})}\subseteq\rightG{(\leftG{\mathcal{P}})}$\quad(\,$\rightleftG{}$ is monotone)
		\item if $\Phi\subseteq\Phi'$ 
				then $\leftG{(\rightG{\Phi})}\subseteq\leftG{(\rightG{\Phi'})}$\quad(\,$\leftrightG{}$ is monotone)
	\end{enumerate}
\end{lemma}
\begin{proof}
	For (1),
		suppose that $\Phi\subseteq\Phi'$.
	Further suppose that $P\in\rightG{\Phi'}$.
	That is,
		$\Phi'\subseteq\Clo{}{}(\invs(P))$.
	Now suppose that $\phi\in\Phi$.
	Hence $\phi\in\Phi'$.
	Hence $\phi\in\Clo{}{}(\invs(P))$.
	Thus $\Phi\subseteq\Clo{}{}(\invs(P))$.
	That is, $P\in\rightG{\Phi}$.
	Thus $\rightG{\Phi'}\subseteq\rightG{\Phi}$.
	For (2),
		suppose that $\mathcal{P}\subseteq\mathcal{P}'$.
	Further suppose that $\phi\in\leftG{\mathcal{P}'}$.
	That is,
		for every $P\in\mathcal{P}'$, $\phi\in\Clo{}{}(\invs(P))$.
	Now suppose that $P\in\mathcal{P}$.
	Hence $P\in\mathcal{P}'$.
	Hence $\phi\in\Clo{}{}(\invs(P))$.
	Thus for every $P\in\mathcal{P}$, $\phi\in\Clo{}{}(\invs(P))$.
	That is, $\phi\in\leftG{\mathcal{P}}$.
	Thus $\leftG{\mathcal{P}'}\subseteq\leftG{\mathcal{P}}$.
	For (3),
		suppose that $P\in\mathcal{P}$.
	Further suppose that $\phi\in\leftG{\mathcal{P}}$.
	That is,
		for every $P\in\mathcal{P}$, $\phi\in\Clo{}{}(\invs(P))$.
	Hence $\phi\in\Clo{}{}(\invs(P))$.
	Thus $\leftG{\mathcal{P}}\subseteq\Clo{}{}(\invs(P))$.
	That is, $P\in\rightG{(\leftG{\mathcal{P}})}$.
	Thus $\mathcal{P}\subseteq\rightG{(\leftG{\mathcal{P}})}$.
	For (4),
		suppose that $\phi\in\Phi$.
	Further suppose that $P\in\rightG{\Phi}$.
	That is,
		$\Phi\subseteq\Clo{}{}(\invs(P))$.
	Hence $\phi\in\Clo{}{}(\invs(P))$.
	Thus for every $P\in\rightG{\Phi}$, $\phi\in\Clo{}{}(\invs(P))$.
	That is, $\phi\in\leftG{(\rightG{\Phi})}$.
	Thus $\Phi\subseteq\leftG{(\rightG{\Phi})}$.
	For (5),
		consider that 
			$\leftG{\mathcal{P}}\subseteq\leftG{(\rightG{(\leftG{\mathcal{P}})})}$ is 
				an instance of (4), and
		that $\leftG{(\rightG{(\leftG{\mathcal{P}})})}\subseteq\leftG{\mathcal{P}}$ by 
			the application of (2) to (3).
	For (6),
		consider that 
			$\rightG{\Phi}\subseteq\rightG{(\leftG{(\rightG{\Phi})})}$ is 
				an instance of (3), and
		that $\rightG{(\leftG{(\rightG{\Phi})})}\subseteq\rightG{\Phi}$ by 
			the application of (1) to (4).
	For (7) and (8),
		substitute $\leftG{P}$ for $\Phi$ in (6), and
		$\rightG{\Phi}$ for $\mathcal{P}$ in (5), respectively.
	For (9) and (10), 
		transitively apply (1) to (2) and (2) to (1), respectively.
\end{proof}
\noindent
Notice that 
	Lemma~\ref{lemma:Properties}.3, \ref{lemma:Properties}.7, and \ref{lemma:Properties}.9 together with 
	Lemma~\ref{lemma:Properties}.4, \ref{lemma:Properties}.8, and \ref{lemma:Properties}.10 mean that 
	$\leftrightG{}$ and $\rightleftG{}$ are closure operators, which is 
		a fact relevant to Theorem~\ref{theorem:DeMorgan}.

\begin{theorem}[The Galois-connection property of $(\,\rightG{}\,,\leftG{}\,)$]\label{theorem:Galois}
	The ordered pair $(\,\rightG{},\leftG{}\,)$ is an \emph{antitone} or \emph{order-reversing Galois-connection} between the powerset algebras in Definition~\ref{definition:PersonalityAlgebras}.
	That is, 
		for every $\Phi\in2^{\statements(\atoms)}$ and $\mathcal{P}\in2^{\SPPs}$,
			$$\text{$\mathcal{P}\subseteq\rightG{\Phi}$ if and only if $\Phi\subseteq\leftG{\mathcal{P}}$.}$$
\end{theorem}
\begin{proof}
	Let $\Phi\in2^{\statements(\atoms)}$ and $\mathcal{P}\in2^{\SPPs}$ and 
	suppose that $\mathcal{P}\subseteq\rightG{\Phi}$.
	Hence $\leftG{(\rightG{\Phi})}\subseteq\leftG{\mathcal{P}}$ by Lemma~\ref{lemma:Properties}.2.
	Further,
		$\Phi\subseteq\leftG{(\rightG{\Phi})}$ by Lemma~\ref{lemma:Properties}.4.
	Hence $\Phi\subseteq\leftG{\mathcal{P}}$ by transitivity.
	Conversely suppose that $\Phi\subseteq\leftG{\mathcal{P}}$.
	Hence $\rightG{(\leftG{\mathcal{P}})}\subseteq\rightG{\Phi}$ by Lemma~\ref{lemma:Properties}.1.
	Further,
		$\mathcal{P}\subseteq\rightG{(\leftG{\mathcal{P}})}$ by Lemma~\ref{lemma:Properties}.3.
	Hence $\mathcal{P}\subseteq\rightG{\Phi}$.
\end{proof}
\noindent
Galois-connections are connected to \emph{residuated mappings} \cite{LatticesAndOrderedAlgebraicStructures}.

\begin{theorem}[De-Morgan like laws]\label{theorem:DeMorgan}\ 
	\begin{enumerate}
		\item 
				$\begin{array}[t]{r@{\ }c@{\ }l@{\ }r@{\ }c@{\ }l@{\ }r@{\ }c@{\ }l@{\ }r@{\ }c@{\ }l}
					\leftG{(\mathcal{P}\cup\mathcal{P}')}&=&
						\leftG{\mathcal{P}}\cap\leftG{\mathcal{P}'}&=&
						\leftG{(\rightG{(\leftG{\mathcal{P}}\cap\leftG{\mathcal{P}'})})}&\subseteq& &&\\
					&&\leftG{\mathcal{P}}\cup\leftG{\mathcal{P}'}&\subseteq&
						\leftG{(\rightG{(\leftG{\mathcal{P}}\cup\leftG{\mathcal{P}'})})}&\subseteq&
						\leftG{(\mathcal{P}\cap\mathcal{P}')}
					\end{array}$
		\item $\begin{array}[t]{r@{\ }c@{\ }l@{\ }r@{\ }c@{\ }l@{\ }r@{\ }c@{\ }l@{\ }r@{\ }c@{\ }l}
					\rightG{(\Phi\cup\Phi')}&=&
						\rightG{\Phi}\cap\rightG{\Phi'}&=&
						\rightG{(\leftG{(\rightG{\Phi}\cap\rightG{\Phi'})})}&\subseteq& &&\\
					&&\rightG{\Phi}\cup\rightG{\Phi'}&\subseteq&
						\rightG{(\leftG{(\rightG{\Phi}\cup\rightG{\Phi'})})}&\subseteq&
						\rightG{(\Phi\cap\Phi')}
				\end{array}$
	\end{enumerate}
\end{theorem}
\begin{proof}
	For $\leftG{(\mathcal{P}\cup\mathcal{P}')}=\leftG{\mathcal{P}}\cap\leftG{\mathcal{P}'}$ (join becomes meet) in (1), 
		let $\phi\in\statements(\atoms)$, and 
		consider that 
			$\phi\in(\leftG{\mathcal{P}\cup\mathcal{P}')}$ if and only if
			(for every $P\in\mathcal{P}\cup\mathcal{P}'$, $\phi\in\Clo{}{}(\invs(P))$) if and only if 
			[for every $P$, ($P\in\mathcal{P}$ or $P\in\mathcal{P}'$) implies $\phi\in\Clo{}{}(\invs(P))$] if and only if 
			[for every $P$, ($P\in\mathcal{P}$ implies $\phi\in\Clo{}{}(\invs(P))$) and 
							($P\in\mathcal{P}'$ implies $\phi\in\Clo{}{}(\invs(P))$)] if and only if 
			[(for every $P\in\mathcal{P}$, $\phi\in\Clo{}{}(\invs(P))$) and 
			 (for every $P\in\mathcal{P}'$, $\phi\in\Clo{}{}(\invs(P))$)] if and only if 
			($\phi\in\leftG{\mathcal{P}}$ and $\phi\in\leftG{\mathcal{P}'}$) if and only if 
			 $\phi\in\leftG{\mathcal{P}}\cap\leftG{\mathcal{P}'}$.
	Then, 
		$\leftG{\mathcal{P}}\cap\leftG{\mathcal{P}'}\subseteq\leftG{\mathcal{P}}\cup\leftG{\mathcal{P}'}$ 
		by elementary set theory.
	For later use of $\leftG{\mathcal{P}}\cup\leftG{\mathcal{P}'}\subseteq\leftG{(\mathcal{P}\cap\mathcal{P}')}$ in (1) consider:
	\begin{enumerate}
		\item\quad $\phi\in\leftG{\mathcal{P}}\cup\leftG{\mathcal{P}'}$\hfill hypothesis\\[-1.5\baselineskip]
		\item\quad $\phi\in\leftG{\mathcal{P}}$ or $\phi\in\leftG{\mathcal{P}'}$\hfill 1\\[-1.5\baselineskip]
		\item\qquad $\phi\in\leftG{\mathcal{P}}$\hfill hypothesis\\[-1.5\baselineskip]
		\item\qquad\quad $P\in\mathcal{P}\cap\mathcal{P}'$\hfill hypothesis\\[-1.5\baselineskip]
		\item\qquad\quad $P\in\mathcal{P}$ and $P\in\mathcal{P}'$\hfill 4\\[-1.5\baselineskip]
		\item\qquad\quad $P\in\mathcal{P}$\hfill 5\\[-1.5\baselineskip]
		\item\qquad\quad $\{P\}\subseteq\mathcal{P}$\hfill 6\\[-1.5\baselineskip]
		\item\qquad\quad $\leftG{\mathcal{P}}\subseteq\leftG{\{P\}}$\hfill 7, Lemma~\ref{lemma:Properties}.2\\[-1.5\baselineskip]
		\item\qquad\quad $\phi\in\leftG{\{P\}}$\hfill 3, 8\\[-1.5\baselineskip]
		\item\qquad\quad $\phi\in\Clo{}{}(\invs(P))$\hfill 9\\[-1.5\baselineskip]
		\item\qquad for every $P\in\mathcal{P}\cap\mathcal{P}'$,
						$\phi\in\Clo{}{}(\invs(P))$\hfill 4--10\\[-1.5\baselineskip]
		\item\qquad $\phi\in\leftG{(\mathcal{P}\cap\mathcal{P}')}$\hfill 11\\[-1.5\baselineskip]
		\item\quad if $\phi\in\leftG{\mathcal{P}}$ then $\phi\in\leftG{(\mathcal{P}\cap\mathcal{P}')}$\hfill 3--12\\[-1.5\baselineskip]
		\item\quad if $\phi\in\leftG{\mathcal{P}'}$ then $\phi\in\leftG{(\mathcal{P}\cap\mathcal{P}')}$\hfill similarly to 3--12 for 13\\[-1.5\baselineskip]
		\item\quad $\phi\in\leftG{(\mathcal{P}\cap\mathcal{P}')}$\hfill 2, 13, 14\\[-1.5\baselineskip]
		\item $\leftG{\mathcal{P}}\cup\leftG{\mathcal{P}'}\subseteq\leftG{(\mathcal{P}\cap\mathcal{P}')}$\hfill 1--15.
	\end{enumerate}
	For $\leftG{(\rightG{(\leftG{\mathcal{P}}\cup\leftG{\mathcal{P}'})})}\subseteq
		\leftG{(\mathcal{P}\cap\mathcal{P}')}$ in (1),
			consider the previously proved property that 
				$\leftG{\mathcal{P}}\cup\leftG{\mathcal{P}'}\subseteq\leftG{(\mathcal{P}\cap\mathcal{P}')}$.
	Hence 	
		$(\mathcal{P}\cap\mathcal{P}')\subseteq\rightG{(\leftG{\mathcal{P}}\cup\leftG{\mathcal{P}'})}$
			by Theorem~\ref{theorem:Galois}.
	Hence 
		$\leftG{(\rightG{(\leftG{\mathcal{P}}\cup\leftG{\mathcal{P}'})})}\subseteq\leftG{(\mathcal{P}\cap\mathcal{P}')}$
			by Lemma~\ref{lemma:Properties}.2.
	Then, 
		$\leftG{\mathcal{P}}\cup\leftG{\mathcal{P}'}\subseteq
			\leftG{(\rightG{(\leftG{\mathcal{P}}\cup\leftG{\mathcal{P}'})})}$ in (1) is an instance of 
				Lemma~\ref{lemma:Properties}.4.
	For $\leftG{\mathcal{P}}\cap\leftG{\mathcal{P}'}=
		\leftG{(\rightG{(\leftG{\mathcal{P}}\cap\leftG{\mathcal{P}'})})}$ in (1),
			consider that 
				$\leftG{(\rightG{(\leftG{\mathcal{P}}\cap\leftG{\mathcal{P}'})})}=
					\leftG{(\rightG{(\leftG{(\mathcal{P}\cup\mathcal{P}')})})}$	by
					 the previously proved property that 
				$\leftG{\mathcal{P}}\cap\leftG{\mathcal{P}'}=\leftG{(\mathcal{P}\cup\mathcal{P}')}$.
	But $\leftG{(\rightG{(\leftG{(\mathcal{P}\cup\mathcal{P}')})})}=
			\leftG{(\mathcal{P}\cup\mathcal{P}')}$ by Lemma~\ref{lemma:Properties}.5.
	Hence $\leftG{(\rightG{(\leftG{\mathcal{P}}\cap\leftG{\mathcal{P}'})})}=
			\leftG{\mathcal{P}}\cap\leftG{\mathcal{P}'}$.
		
	For $\rightG{(\Phi\cup\Phi')}=\rightG{\Phi}\cap\rightG{\Phi'}$ (join becomes meet) in (2), 
		let $P\in\SPPs$, and 
		consider that 
			$P\in(\rightG{\Phi\cup\Phi')}$ if and only if
			(for every $\phi\in\Phi\cup\Phi'$, $\phi\in\Clo{}{}(\invs(P))$) if and only if 
			[for every $\phi$, ($\phi\in\Phi$ or $\phi\in\Phi'$) implies $\phi\in\Clo{}{}(\invs(P))$] if and only if 
			[for every $\phi$, 
				($\phi\in\Phi$ implies $\phi\in\Clo{}{}(\invs(P))$) and 
				($\phi\in\Phi'$ implies $\phi\in\Clo{}{}(\invs(P))$)] if and only if 
			[(for every $\phi\in\Phi$, $\phi\in\Clo{}{}(\invs(P))$) and 
			 (for every $\phi\in\Phi'$, $\phi\in\Clo{}{}(\invs(P))$)] if and only if 
			($P\in\rightG{\Phi}$ and $P\in\rightG{\Phi'}$) if and only if 
			 $P\in\rightG{\Phi}\cap\rightG{\Phi'}$.
	Then, 
		$\rightG{\Phi}\cap\rightG{\Phi'}\subseteq\rightG{\Phi}\cup\rightG{\Phi'}$ 
		by elementary set theory.
	For later use of $\rightG{\Phi}\cup\rightG{\Phi'}\subseteq\rightG{(\Phi\cap\Phi')}$ in (2) consider: 
	\begin{enumerate}
		\item\quad $P\in\rightG{\Phi}\cup\rightG{\Phi'}$\hfill hypothesis\\[-1.5\baselineskip]
		\item\quad $P\in\rightG{\Phi}$ or $P\in\rightG{\Phi'}$\hfill 1\\[-1.5\baselineskip]
		\item\qquad $P\in\rightG{\Phi}$\hfill hypothesis\\[-1.5\baselineskip]
		\item\qquad\quad $\phi\in\Phi\cap\Phi'$\hfill hypothesis\\[-1.5\baselineskip]
		\item\qquad\quad $\phi\in\Phi$ and $\phi\in\Phi'$\hfill 4\\[-1.5\baselineskip]
		\item\qquad\quad $\phi\in\Phi$\hfill 5\\[-1.5\baselineskip]
		\item\qquad\quad $\{\phi\}\subseteq\Phi$\hfill 6\\[-1.5\baselineskip]
		\item\qquad\quad $\rightG{\Phi}\subseteq\rightG{\{\phi\}}$\hfill 7, Lemma~\ref{lemma:Properties}.1\\[-1.5\baselineskip]
		\item\qquad\quad $P\in\rightG{\{\phi\}}$\hfill 3, 8\\[-1.5\baselineskip]
		\item\qquad\quad $\phi\in\Clo{}{}(\invs(P))$\hfill 9\\[-1.5\baselineskip]
		\item\qquad for every $\phi\in\Phi\cap\Phi'$,
						$\phi\in\Clo{}{}(\invs(P))$\hfill 4--10\\[-1.5\baselineskip]
		\item\qquad $P\in\rightG{(\Phi\cap\Phi')}$\hfill 11\\[-1.5\baselineskip]
		\item\quad if $P\in\rightG{\Phi}$ then $P\in\rightG{(\Phi\cap\Phi')}$\hfill 3--12\\[-1.5\baselineskip]
		\item\quad if $P\in\rightG{\Phi'}$ then $P\in\rightG{(\Phi\cap\Phi')}$\hfill similarly to 3--12 for 13\\[-1.5\baselineskip]
		\item\quad $P\in\rightG{(\Phi\cap\Phi')}$\hfill 2, 13, 14\\[-1.5\baselineskip]
		\item $\rightG{\Phi}\cup\rightG{\Phi'}\subseteq\rightG{(\Phi\cap\Phi')}$\hfill 1--15.
	\end{enumerate}
	For $\rightG{(\leftG{(\rightG{\Phi}\cup\rightG{\Phi'})})}\subseteq
		\rightG{(\Phi\cap\Phi')}$ in (2),
			consider the previously proved property that 
				$\rightG{\Phi}\cup\rightG{\Phi'}\subseteq\rightG{(\Phi\cap\Phi')}$.
	Hence 	
		$(\Phi\cap\Phi')\subseteq\leftG{(\rightG{\Phi}\cup\rightG{\Phi'})}$
			by Theorem~\ref{theorem:Galois}.
	Hence 
		$\rightG{(\leftG{(\rightG{\Phi}\cup\rightG{\Phi'})})}\subseteq\rightG{(\Phi\cap\Phi')}$
			by Lemma~\ref{lemma:Properties}.1.
	Then, 
		$\rightG{\Phi}\cup\rightG{\Phi'}\subseteq
			\rightG{(\leftG{(\rightG{\Phi}\cup\rightG{\Phi'})})}$ in (2) is an instance of 
				Lemma~\ref{lemma:Properties}.3.
	For $\rightG{\Phi}\cap\rightG{\Phi'}=
		\rightG{(\leftG{(\rightG{\Phi}\cap\rightG{\Phi'})})}$ in (2),
			consider that 
				$\rightG{(\leftG{(\rightG{\Phi}\cap\rightG{\Phi'})})}=
					\rightG{(\leftG{(\rightG{(\Phi\cup\Phi')})})}$	by
					 the previously proved property that 
				$\rightG{\Phi}\cap\rightG{\Phi'}=\rightG{(\Phi\cup\Phi')}$.
	But $\rightG{(\leftG{(\rightG{(\Phi\cup\Phi')})})}=
			\rightG{(\Phi\cup\Phi')}$ by Lemma~\ref{lemma:Properties}.6.
	Hence $\rightG{(\leftG{(\rightG{\Phi}\cap\rightG{\Phi'})})}=
			\rightG{\Phi}\cap\rightG{\Phi'}$.
\end{proof}

\begin{corollary}\label{corollary:WellDefinedness}
	The quotient algebras in Table~\ref{table:LTAlgebras} are well-defined, that is,
			the equivalence relations 
		${\equiv}\subseteq2^{\statements(\atoms)}\times2^{\statements(\atoms)}$ and 
		${\equiv}\subseteq2^{\SPPs}\times2^{\SPPs}$ are \emph{congruences:}
		\begin{enumerate}
			\item if 
					$\Phi\equiv\Phi'$ and 
					$\Phi''\equiv\Phi'''$
				then 
					$\Phi\cup\Phi''\equiv\Phi'\cup\Phi'''$;
			\item if 
					$\mathcal{P}\equiv\mathcal{P}'$ and 
					$\mathcal{P}''\equiv\mathcal{P}'''$
				then 
					$\mathcal{P}\cup\mathcal{P}''\equiv\mathcal{P}'\cup\mathcal{P}'''$.
		\end{enumerate}
\end{corollary}
\begin{proof}
	By the De-Morgan like laws  
		$\rightG{(\Phi\cup\Phi')}=\rightG{\Phi}\cap\rightG{\Phi'}$ and 
		$\leftG{(\mathcal{P}\cup\mathcal{P}')}=\leftG{\mathcal{P}}\cap\leftG{\mathcal{P}'}$, respectively 
			(see Theorem~\ref{theorem:DeMorgan}).
\end{proof}

We are finally ready for defining our announced personality categories, and this 
	by means of our previously-defined Galois-connection.
\begin{definition}[Personality categories]
Let 
	$\mathcal{P}\in2^{\SPPs}$ and 
	$\Phi\in2^{\statements(\atoms)}$, and
let  
	\begin{itemize}
		\item $\mathcal{T}_{\Phi}:=\{\,\tau:2^{\SPPs}\to2^{\SPPs}\mid
				\text{\begin{tabular}{@{\ }c@{\ }}
						for every $\mathcal{P}\in2^{\SPPs}$ and $\phi\in\Phi$,\\  
						$\phi\in\leftG{\mathcal{P}}$ implies $\phi\in\leftG{\tau(\mathcal{P})}$
					 \end{tabular}}\,\}$ and 
		\item $\mathcal{T}_{\mathcal{P}}:=\{\,\tau:2^{\statements(\atoms)}\to2^{\statements(\atoms)}\mid
				\text{\begin{tabular}{@{\ }c@{\ }}
						for every $\Phi\in2^{\statements(\atoms)}$ and $P\in\mathcal{P}$,\\  
						$P\in\rightG{\Phi}$ implies $P\in\rightG{\tau(\Phi)}$
					 \end{tabular}}\,\}$\,. 
	\end{itemize}
	Then, define the categories (monoids) 
		$$\text{$\mathbf{T}_{\Phi}:=\langle\mathcal{T}_{\Phi},\id,\circ\rangle$ and 
		$\mathbf{T}_{\mathcal{P}}:=\langle\mathcal{T}_{\mathcal{P}},\id,\circ\rangle$}$$ of 
			$\Phi$- and $\mathcal{P}$-preserving transformations, respectively.
\end{definition}

\begin{proposition}[Antitonicity properties of personality categories]\label{proposition:GPPC}\ 
	\begin{enumerate}
		\item $\Phi\subseteq\Phi'$ implies $\mathcal{T}_{\Phi'}\subseteq\mathcal{T}_{\Phi}$
		\item $\mathcal{P}\subseteq\mathcal{P}'$ implies $\mathcal{T}_{\mathcal{P}'}\subseteq\mathcal{T}_{\mathcal{P}}$
		\item $P\sqsubseteq P'$ implies $\mathcal{T}_{\leftG{\{P'\}}}\subseteq\mathcal{T}_{\leftG{\{P\}}}$
		\item $\mathcal{T}_{\Phi\cup\Phi'}\subseteq\mathcal{T}_{\Phi}\subseteq\mathcal{T}_{\Phi\cap\Phi'}$
		\item $\mathcal{T}_{\mathcal{P}\cup\mathcal{P}'}\subseteq\mathcal{T}_{\mathcal{P}}\subseteq\mathcal{T}_{\mathcal{P}\cap\mathcal{P}'}$
	\end{enumerate}
\end{proposition}
\begin{proof}
	(1) and (2) follow straightforwardly from their respective definition, and 
	(4) and (5) from (1) and (2), respectively. 
	(3) follows from 
			Proposition~\ref{proposition:SpecialCase}.2 and 
			(1) by transitivity.
\end{proof}

\begin{proposition}[Preservation properties of personality transformations]\ 
	\begin{enumerate}
		\item $\tau\in\mathcal{T}_{\rightG{\Phi}}$ implies $\rightG{\Phi}\subseteq\rightG{\tau(\Phi)}$
		\item $\tau\in\mathcal{T}_{\leftG{\mathcal{P}}}$ implies $\leftG{\mathcal{P}}\subseteq\leftG{\tau(\mathcal{P})}$
		\item $\tau\in\mathcal{T}_{\rightG{(\Phi\cap\Phi')}}$ implies 
				($\rightG{\Phi}\subseteq\rightG{\tau(\Phi)}$ and $\rightG{\Phi'}\subseteq\rightG{\tau(\Phi')}$)
		\item $\tau\in\mathcal{T}_{\leftG{(\mathcal{P}\cap\mathcal{P}')}}$ implies 
				($\leftG{\mathcal{P}}\subseteq\leftG{\tau(\mathcal{P})}$ and $\leftG{\mathcal{P}'}\subseteq\leftG{\tau(\mathcal{P}')}$)
	\end{enumerate}
\end{proposition}
\begin{proof}
	(1) and (2) follow by expansion of definitions.
	For (3) suppose that $\tau\in\mathcal{T}_{\rightG{(\Phi\cap\Phi')}}$.
	But by Theorem~\ref{theorem:DeMorgan}.2,
		$\rightG{\Phi}\cup\rightG{\Phi'}\subseteq\rightG{(\Phi\cap\Phi')}$.
	Hence $\mathcal{T}_{\rightG{(\Phi\cap\Phi')}}\subseteq\mathcal{T}_{\rightG{\Phi}\cup\rightG{\Phi'}}$ by
		Proposition~\ref{proposition:GPPC}.1.
	Hence $\tau\in\mathcal{T}_{\rightG{\Phi}\cup\rightG{\Phi'}}$.
	Hence $\tau\in\mathcal{T}_{\rightG{\Phi}}$ and $\tau\in\mathcal{T}_{\rightG{\Phi'}}$ by
		Proposition~\ref{proposition:GPPC}.4.
	Hence $\rightG{\Phi}\subseteq\rightG{\tau(\Phi)}$ and $\rightG{\Phi'}\subseteq\rightG{\tau(\Phi')}$ by (1).
	For (4) suppose that $\tau\in\mathcal{T}_{\leftG{(\mathcal{P}\cap\mathcal{P}')}}$.
	But by Theorem~\ref{theorem:DeMorgan}.1,
		$\leftG{\mathcal{P}}\cup\leftG{\mathcal{P}'}\subseteq\leftG{(\mathcal{P}\cap\mathcal{P}')}$.
	Hence $\mathcal{T}_{\leftG{(\mathcal{P}\cap\mathcal{P}')}}\subseteq\mathcal{T}_{\leftG{\mathcal{P}}\cup\leftG{\mathcal{P}'}}$ by
		Proposition~\ref{proposition:GPPC}.2.
	Hence $\tau\in\mathcal{T}_{\leftG{\mathcal{P}}\cup\leftG{\mathcal{P}'}}$.
	Hence $\tau\in\mathcal{T}_{\leftG{\mathcal{P}}}$ and $\tau\in\mathcal{T}_{\leftG{\mathcal{P}'}}$ by
		Proposition~\ref{proposition:GPPC}.5.
	Hence $\leftG{\mathcal{P}}\subseteq\leftG{\tau(\mathcal{P})}$ and $\leftG{\mathcal{P}'}\subseteq\leftG{\tau(\mathcal{P}')}$ by (2).
\end{proof}

\section{Conclusions}
We have provided a formal framework for 
	the computer-aided discovery and categorisation of personality axioms as 
		summarised in the abstract of this paper.
Our framework is meant as 
	a contribution towards practicing
		psychological research with the methods of the exact sciences, for obvious ethical reasons.
Psychology workers (psychologists, psychiatrists, etc.) can now apply 
	our visual framework in their own field (of) studies in order to 
		discover personality theories and categories of their own interest.
Our hope is that these field studies will lead to
	a mathematical systematisation of 
		the academic discipline of psychology in the area of test-based personality theories  
			with the help of our framework.

As future work on our current \emph{synchronic} data analytics approach, which 
	infers \emph{perfect} implicational correlations (between human reactions) at 
		a given time point (within a Szondi personality profile, an SPP) from 
			their invariance across time (within an SPP-sequence, a Szondi-test result), 
				\emph{approximate} implicational correlations can be studied and 
					a \emph{diachronic} data analytics approach can be taken.
Actually, 
	our implicational diagrams such as Table~\ref{table:implicationalVGPSzondi} already contain 
		such approximate implicational correlations in the form of cell values greater than 0, which 
			as explained on Page~\pageref{page:DistanceToInvariance} indicate the distance to 
				invariance and thus the approximation to the perfection in question.
This notion of approximate implicational correlation can be 
	understood and further studied as a notion of \emph{fuzzy implication} \cite{FuzzyImplications}.
Then, a diachronic approach would mine correlations between 
	\emph{different} time points, typically one or several past or present and one or several future, 
		in order to \emph{forecast and predict} future reactions of the person in question, such
			as can be done with \emph{Bayesian inference} \cite{Causality} and \emph{time series} analysis and forecasting \cite{ChaosTimeSeries}.
Actually,
	Table~\ref{table:VGPSzondi} is such a time series.
	
Last but not least,
	we mention the only piece of related work \cite{InvariantRelationsInAFiniteDomain} that
		we are aware of.
There,
	the author develops a framework similarly motivated by invariance as ours, but 
		with quite different setup, outcomes, and results.
The author's setup on the invariants side is 
	a set of relations over 
		a finite domain closed under the Boolean operations, whereas
			our corresponding setup is 
				an intuitionistic theory, a certain set of propositional formulas, as 
					induced by a data sequence (as exemplified by  
						one produced by a personality test).
On the transformations side,
	the author's setup is a system of 
		injective total functions, whereas
			our corresponding setup is a system of total functions \emph{tout court}.

\paragraph{Acknowledgements} 
%
The \LaTeX-packages Listings and TikZ were helpful.

\bibliographystyle{plain}

\begin{thebibliography}{10}

\bibitem{FoundationsOfDatabases}
S.~Abiteboul, R.~Hull, and V.~Vianu.
\newblock {\em Foundations of Databases}.
\newblock Addison-Wesley, 1995.

\bibitem{LatticesAndOrderedAlgebraicStructures}
T.S. Blyth.
\newblock {\em Lattices and Ordered Algebraic Structures}.
\newblock Springer, 2005.

\bibitem{InvariantRelationsInAFiniteDomain}
L.~Burigana.
\newblock Invariant relations in a finite domain.
\newblock {\em Math{\'e}matiques et sciences humaines}, 169, 2005.

\bibitem{DaveyPriestley}
B.A. Davey and H.A. Priestley.
\newblock {\em Introduction to Lattices and Order}.
\newblock Cambridge University Press, 2nd edition, 1990 (2002).

\bibitem{DeJonghOnDP}
D.~de~Jongh.
\newblock The disjunction property according to \mbox{Kreisel} and
  \mbox{Putnam}, 2009.

\bibitem{LoopInvariants}
C.A. Furia, B.~Meyer, and S.~Velder.
\newblock Loop invariants: Analysis, classification, and examples.
\newblock {\em {ACM} Computing Surveys}, 46(3), 2014.

\bibitem{WhatIsALogicalSystem}
D.M. Gabbay, editor.
\newblock {\em What Is a Logical System?}
\newblock Number~4 in Studies in Logic and Computation. Oxford University
  Press, 1995.

\bibitem{Goedel:OnIPC}
K.~G{\"o}del.
\newblock {\em Kurt G{\"o}del: Collected Works}, volume~I, chapter On the
  intuitionistic propositional calculus.
\newblock Oxford University Press, 1986.

\bibitem{ErlangerProgramm}
T.~Hawkings.
\newblock The {Erlanger} programm of {Felix} {Klein:} reflections on its place
  in the history of mathematics.
\newblock {\em Historia Mathematica}, 11, 1984.

\bibitem{FuzzyImplications}
B.~Jayaram and M.~Baczy\'{n}ski.
\newblock {\em Fuzzy Implications}.
\newblock Springer, 2008.

\bibitem{Klein:ErlangerProgramm}
F.~Klein.
\newblock Vergleichende {B}etrachtungen {\"u}ber neuere geometrische
  {F}orschungen.
\newblock {\em Reprinted with additional footnotes in Mathematische Annalen},
  43 (1893), 1872.

\bibitem{Klein:Geometry}
F.~Klein.
\newblock {\em Elementary Mathematics from an Advanced Standpoint: Geometry}.
\newblock Dover Publications, 2004.

\bibitem{arXiv:1403.2000v1}
S.~Kramer.
\newblock A {Galois}-connection between {Myers}-{Briggs'} {Type} {I}ndicators
  and {Szondi's} {P}ersonality {P}rofiles.
\newblock Technical Report 1403.2000, arXiv, 2014.
\newblock \url{http://arxiv.org/abs/1403.2000}.

\bibitem{SzondiTestWebApp}
S.~Kramer.
\newblock \url{www.szondi-test.ch}, 2014.
\newblock forthcoming.

\bibitem{KripkeSemanticsIL}
S.A. Kripke.
\newblock {\em Formal Systems and Recursive Functions}, volume~40 of {\em
  Studies in Logic and the Foundations of Mathematics}, chapter Semantical
  Analysis of Intuitionistic Logic I.
\newblock Elsevier, 1965.

\bibitem{sep-category-theory}
Jean-Pierre Marquis.
\newblock Category theory.
\newblock In Edward~N. Zalta, editor, {\em The Stanford Encyclopedia of
  Philosophy}. Summer 2013 edition, 2013.

\bibitem{sep-logic-intuitionistic}
J.~Moschovakis.
\newblock Intuitionistic logic.
\newblock In Edward~N. Zalta, editor, {\em The Stanford Encyclopedia of
  Philosophy}. Summer 2010 edition, 2010.

\bibitem{Causality}
J.~Pearl.
\newblock {\em Causality}.
\newblock Cambridge University Press, 2nd edition, 2009.

\bibitem{ChaosTimeSeries}
J.C. Sprott.
\newblock {\em Chaos and Time-Series Analysis}.
\newblock Oxford University Press, 2003.

\bibitem{Szondi:Triebpathologie:IA}
L.~Szondi.
\newblock {\em Triebpathologie. Teil A: Dialektische Trieblehre und
  dialektische Methodik der Testanalyse}, volume~1.
\newblock Hans Huber, 2nd edition, 1952 (1977).

\bibitem{Szondi:ETD:Band1}
L.~Szondi.
\newblock {\em Lehrbuch der Experimentellen Triebdiagnostik}, volume I:
  Text-Band.
\newblock Hans Huber, 3rd edition, 1972.

\bibitem{Szondi:IchAnalyse}
L.~Szondi.
\newblock {\em Ich-Analyse: Die Grundlage zur Vereinigung der
  Tiefenpsychologie}.
\newblock Hans Huber, 1999.
\newblock English translation:
  \url{https://sites.google.com/site/ajohnstontranslationsofszondi/}.

\end{thebibliography}

\end{document}